\documentclass[aps,prl,twocolumn,superscriptaddress]{revtex4-2}
\usepackage{amsmath,amssymb,graphicx,color,bm,epstopdf,comment,braket}
\usepackage[colorlinks=true,allcolors=blue]{hyperref}

\begin{document}

\title{Rydberg-mediated microwave-to-optical transduction with tunable bandwidth}

\author{Iason Tsiamis}
\email{iason.tsiamis@fu-berlin.de}
\affiliation{Dahlem Center for Complex Quantum Systems and Fachbereich Physik, Freie Universität Berlin, 14195 Berlin, Germany}

\author{David Petrosyan}
\affiliation{Institute of Electronic Structure and Laser and Center for Quantum Science and Technologies, FORTH, 70013 Heraklion, Crete, Greece}
\affiliation{Center for Quantum Science, Physikalisches Institut, Eberhard Karls Universität Tübingen, Auf der Morgenstelle 14, 72076 Tübingen, Germany}
\author{Anders S. S{\o}rensen}
\affiliation{Center for Hybrid Quantum Networks (Hy-Q), The Niels Bohr Institute, University of Copenhagen, DK-2100 Copenhagen {\O}, Denmark}

\begin{abstract}
Coherent microwave-to-optical transduction is essential for hybrid quantum networks. Typical transducers, however, have bandwidths set by fixed device parameters 
and face efficiency--bandwidth tradeoffs. We propose a cavity-based microwave-to-optical transducer mediated by a collective Rydberg excitation, where external driving fields tune the conversion bandwidth \textit{in situ} from narrowband, spectrally selective operation to broadband operation without modifying cavity parameters. Under  impedance-matching conditions, the system operates as a continuous-wave  phase-preserving linear transducer, maintaining both near-unit efficiency and  quantum-limited added noise throughout the tunable bandwidth range.
\end{abstract}

\maketitle

Hybrid quantum networks require coherent interfaces between microwave-frequency quantum processors and optical communication channels~\cite{Kimble2008,Lauk2020}. Superconducting circuits, in particular, provide a versatile platform for local
quantum information processing~\cite{Blais2021}, but microwave photons---although
coherently transmitted over meter-scale cryogenic links~\cite{Magnard2020}---suffer
prohibitive transmission loss and thermal decoherence over network-scale distances. Optical photons, by contrast, propagate with minimal loss in fibers and are natural carriers for distributed quantum communication~\cite{Kimble2008,Wehner2018}. Coherent, efficient, and low-noise microwave-to-optical transduction would
therefore enable superconducting processors to interface with optical channels
for quantum communication~\cite{Lauk2020,Kurizki2015,Clerk2020}.

A prominent approach to microwave-to-optical transduction \cite{Zeuthen2020} utilizes cavity optomechanics, where a mechanical mode mediates interactions between microwave and optical fields~\cite{Andrews2014,Higginbotham2018,Barzanjeh2012}. While providing a clear input-output framework for efficient conversion, mechanical mediators introduce thermal noise, have limited bandwidth, and are sensitive to optical heating~\cite{Higginbotham2018,Lauk2020}. Electro-optic and integrated photonic approaches avoid mechanical intermediaries, yet they remain constrained by weak nonlinearities, demanding microwave--optical mode matching, and highly demanding cavity parameters~\cite{Rueda2016,JaverzacGaly2016,Mirhosseini2020}.

Rydberg atomic ensembles offer a complementary route to microwave--optical transduction, combining large microwave transition dipole moments with optical transitions in the same medium~\cite{Petrosyan2019,Covey2019}. Previous free-space Rydberg experiments have demonstrated microwave-to-optical conversion via multiwave mixing and electromagnetically induced transparency~\cite{Han2018,Vogt2019,Tu2022,Borowka2023}. The latter experiments achieve high conversion efficiencies and continuous-wave operation~\cite{Tu2022,Borowka2023}, a regime also explored for single-photon-level control in microwave and optical platforms~\cite{Kyriienko2016,TsiamisSPT2026,TsiamisDephasing2026}. While free-space implementations emit into a continuum of optical modes,
cavity-assisted Rydberg architectures can enable impedance matching through
cavity-defined channels~\cite{Kumar2023}. In the demonstrated cavity-EIT
implementation, however, the conversion remains narrowband.

A common limitation of previous microwave-to-optical transducers is that the conversion bandwidth is tied to effective linewidths and coupling rates, leading to an efficiency--bandwidth tradeoff set by fixed device parameters. Different applications, however, favor different spectral responses: narrowband conversion can act as a spectral filter, suppressing out-of-band background fields~\cite{Delaney2022,Mirhosseini2020}, whereas broadband conversion can accommodate shorter wave packets and higher operation rates~\cite{Sahu2022,Arnold2025}.
 The ability to reconfigure the conversion bandwidth \textit{in situ}, while maintaining high conversion efficiency, would therefore allow a transducer to adapt to different sources and network tasks without modifying the device hardware.

In this Letter, we propose a cavity-based Rydberg-mediated scheme for 
continuous-wave microwave-to-optical transduction that realizes this 
capability. Microwave and optical cavity modes collectively couple to the atomic ensemble, forming a phase-preserving linear transducer. Under impedance-matching conditions, the conversion efficiency approaches unity with quantum-limited added noise. The conversion bandwidth is continuously tunable \textit{in situ} through external driving fields, allowing the same transducer to span operating regimes from narrowband, spectrally selective conversion to broadband conversion without modifying cavity parameters.

We analyze the transducer for both equal and unequal microwave and optical
cavity linewidths, showing that efficient broadband conversion and tunable spectral control remain robust against experimentally relevant nonuniform atom--cavity coupling and dissipation. While the framework applies broadly to cavity-based Rydberg transduction platforms, we illustrate its experimental feasibility by evaluating a representative atom--cavity implementation and the corresponding parameter requirements.

\begin{figure}%[h!]
\includegraphics[width=1.\columnwidth]{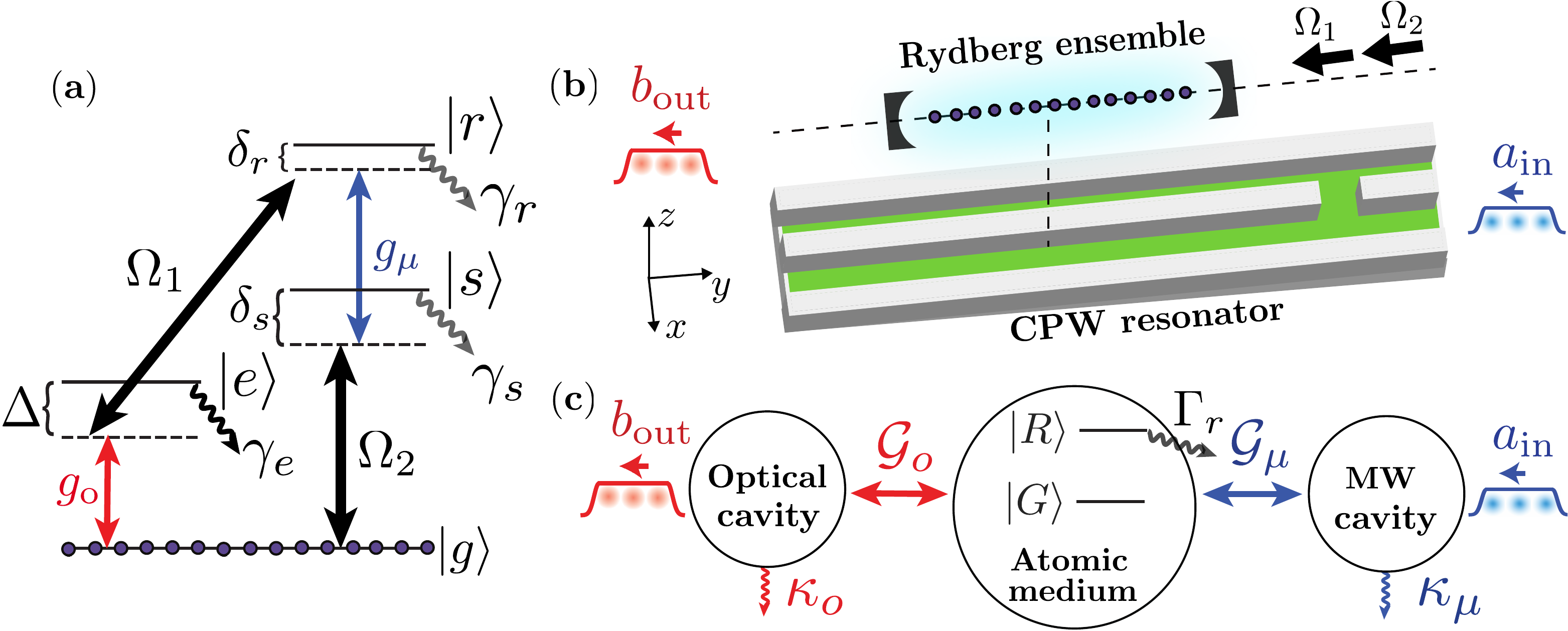}
\caption{
(a) Atomic level structure and coupling scheme. A microwave cavity mode couples the Rydberg transition $\ket{s}\!\leftrightarrow\!\ket{r}$ with strength $g_\mu$, while an optical cavity mode couples $\ket{g}\!\leftrightarrow\!\ket{e}$ with strength $g_o$. Two classical control fields with Rabi frequencies $\Omega_1$ and $\Omega_2$ drive the transitions $\ket{r}\!\leftrightarrow\!\ket{e}$ and $\ket{s}\!\leftrightarrow\!\ket{g}$, respectively. Detunings and decay rates are indicated.
(b) Possible experimental realization. An ensemble of Rydberg atoms is trapped near a coplanar-waveguide microwave resonator and coupled to an optical cavity mode. The collective atomic medium mediates coherent conversion between microwave and optical fields.
(c) Effective three-mode description. In the dispersive regime, the off-resonant states $\ket{e}$ and $\ket{s}$ are adiabatically eliminated, yielding effective Raman couplings $\mathcal G_o$ and $\mathcal G_\mu$ between the collective spin-wave mode $R$ and the optical and microwave cavity modes, respectively. The system acts as a linear transducer between the input microwave field $a_{\rm in}$ and the output optical field $b_{\rm out}$.}
\label{fig:sketch}
\end{figure}

\textit{System.---}
We consider an ensemble of $N$ atoms with four relevant internal states: the ground state $\ket{g}$, the excited state $\ket{e}$, and two Rydberg states $\ket{s}$ and $\ket{r}$, see Fig.~\ref{fig:sketch}(a). A microwave cavity mode of frequency $\omega_\mu$ and linewidth $\kappa_\mu$ couples the Rydberg transition $\ket{s}\!\leftrightarrow\!\ket{r}$ with coupling strength $g_\mu$. An optical cavity mode of frequency $\omega_o$ and linewidth $\kappa_o$ couples the transition $\ket{g}\!\leftrightarrow\!\ket{e}$ with coupling strength $g_o$. For simplicity, we first consider uniform, real atom-cavity couplings. Two driving fields with real Rabi frequencies $\Omega_1$ and $\Omega_2$, and frequencies $\omega_{1,2}$, drive the transitions $\ket{e}\!\leftrightarrow\!\ket{r}$ and $\ket{s}\!\leftrightarrow\!\ket{g}$, respectively. In a frame rotating with the cavity and driving-field frequencies, and under the four-photon resonance condition $\omega_\mu+\omega_2=\omega_o+\omega_1$, the Hamiltonian reads
\begin{align}
\frac{\tilde H}{\hbar}
={}&
\sum_{l=1}^N
\left(
\delta_s \sigma_{ss}^{(l)}
+\delta_r \sigma_{rr}^{(l)}
+\Delta \sigma_{ee}^{(l)}
\right) \notag\\
&+\sum_{l=1}^N \Big[
g_\mu a^\dagger \sigma_{sr}^{(l)}
+g_o b^\dagger e^{-i\mathbf K\cdot\mathbf r_l}\sigma_{ge}^{(l)}
\notag\\
&\qquad\quad
+\Omega_1 \sigma_{re}^{(l)}
+\Omega_2 \sigma_{sg}^{(l)}
+\mathrm{H.c.}
\Big],
\label{eq:H_system}
\end{align}
where $a$ ($b$) annihilates a microwave (optical) cavity photon, and $\sigma_{\alpha\beta}^{(l)}=\ket{\alpha}_l\!\bra{\beta}$ is the transition operator between states $\ket{\beta}$ and $\ket{\alpha}$ of the $l$th atom. The atomic position is denoted by $\mathbf r_l$, and $\mathbf K=\mathbf k_o+\mathbf k_1-\mathbf k_2$ is the net phase-mismatch wavevector, where $\mathbf k_o$ and $\mathbf k_{1,2}$ denote the wavevectors of the travelling-wave optical-cavity field and the two driving fields, respectively. For a standing-wave cavity, the corresponding atom--cavity interaction has the
same form for atoms localized near a single antinode, with no cavity contribution to $\mathbf K$; for ensembles spanning multiple antinodes, the oscillating standing-wave mode profile must be retained explicitly. The detunings are $\Delta = \omega_e-\omega_o$, $\delta_s = \omega_s-\omega_2$, and $\delta_r = \omega_r-(\omega_o+\omega_1)$. The spontaneous emission rates of the states $\ket{e}$, $\ket{s}$, and $\ket{r}$ are denoted by $\gamma_{e,s,r}$, where the Rydberg-state decay rates satisfy $\gamma_{s,r}\ll\gamma_e$. We neglect Rydberg--Rydberg interactions for sufficiently weak input fields and/or low-density atomic ensembles.

Although the framework developed below is general and applies to a broad class of microwave--optical cavity systems coupled to Rydberg ensembles, we briefly outline a possible experimental implementation. An ensemble of Rydberg atoms is trapped near a coplanar waveguide resonator, allowing nearly uniform coupling to the microwave cavity mode~\cite{Tsiamis2026,Kaiser2022}. The ensemble is simultaneously coupled to an optical cavity, as illustrated in Fig.~\ref{fig:sketch}(b). An input microwave field incident on the microwave resonator is converted, via a collective Rydberg excitation, into an optical output field from the optical cavity.

We operate in the far-off-resonant regime,
$
|\Delta|
\gg
|\Omega_1|,\,
|g_o| \sqrt{N\,\langle b^\dagger b + 1\rangle}
$
and
$
|\delta_s|
\gg|\Omega_2|\sqrt{N},\,
|g_\mu| \sqrt{\langle a^\dagger a + 1\rangle}.
$
Under these conditions, the states $\ket{e}$ and $\ket{s}$ are only virtually populated and can be adiabatically eliminated~\cite{Reiter2012}. Furthermore, in the weak-excitation limit
$
\langle \sigma_{gg}^{(l)} \rangle \approx 1,
$
the dynamics reduce to those of the collective spin-wave mode described by the operator
$
R=\frac{1}{\sqrt N}\sum_l e^{-i\mathbf K\cdot\mathbf r_l}\sigma_{gr}^{(l)},
$
coupled to the cavity modes through collectively enhanced Raman couplings
$
\mathcal G_\mu=-\sqrt N\,g_\mu\Omega_2/\delta_s
$
and
$
\mathcal G_o=-\sqrt N\,g_o\Omega_1/\Delta$, provided that $\Delta \gg \gamma_e$ and $\delta_s \gg \gamma_s$.

The resulting effective Hamiltonian, under near-phase-matching conditions $\mathbf K\approx0$, describes a three-mode system
\begin{equation}
\frac{H_{\mathrm{eff}}}{\hbar}
=
\delta_{\rm eff} R^\dagger R
+
\left(
\mathcal G_\mu a^\dagger R
+
\mathcal G_o b^\dagger R
+ \mathrm{H.c.}
\right),
\label{eq:Heff}
\end{equation}
where $\delta_{\rm eff}$ is the effective detuning including the induced ac Stark shifts. The resonance condition $\delta_{\rm eff}\approx0$ is obtained by tuning $\delta_r$ to compensate the induced shifts. Dissipative processes associated with the eliminated states lead to an effective Rydberg-state linewidth $\Gamma_r
\approx
\gamma_r
+
\gamma_e \Omega_1^2/\Delta^2$, where the second term is the leading Raman-induced contribution (see Supplemental Material). The effective model is illustrated in Fig.~\ref{fig:sketch}(c).

\textit{Impedance matching.---}%
Including the cavity linewidths $\kappa_\mu$, $\kappa_o$ and the effective Rydberg linewidth $\Gamma_r$, the Heisenberg-Langevin equations derived from Eq.~(\ref{eq:Heff}), together with the corresponding input-output relations, yield the coherent microwave-to-optical conversion amplitude $S_{o\mu}(\omega)$, defined by $\langle b_{\rm out}(\omega)\rangle=S_{o\mu}(\omega)\langle a_{\rm in}(\omega)\rangle$ in the absence of coherent optical and Rydberg-channel inputs. It is given by 
\begin{equation}
S_{o\mu}(\omega)=
-i\sqrt{\kappa_\mu\kappa_o}\,
\frac{\mathcal G_\mu\mathcal G_o}{\mathcal{D}(\omega)},
\label{eq:Somega}
\end{equation}
where $\omega$ denotes the frequency offset from the cavity resonances and $\mathcal{D}(\omega)=\chi_\mu^{-1}\chi_o^{-1}\chi_r^{-1}-\chi_\mu^{-1}\mathcal G_o^2-\chi_o^{-1}\mathcal G_\mu^2$ is the linear-response denominator, with $\chi_{\mu,o}^{-1}=\omega+i\kappa_{\mu,o}/2$ and $\chi_r^{-1}=\omega-\delta_{\rm eff}+i\Gamma_r/2$. The conversion efficiency is $\eta(\omega)=|S_{o\mu}(\omega)|^2$. On resonance ($\omega=0$), the peak efficiency becomes
\begin{equation}
\eta(0)=
\frac{4\Gamma_\mu\Gamma_o}{(\Gamma_r+\Gamma_\mu+\Gamma_o)^2},
\label{eq:heta}
\end{equation}
where $\Gamma_{j}=4\mathcal G_{j}^2/\kappa_{j}$, with $j=\mu,o$, are the effective cavity-induced decay rates of the collective spin-wave mode into the microwave and optical channels. The conversion is maximized under the impedance-matching condition $\Gamma_\mu=\Gamma_o$~\cite{Pinotsi2008}, for which the collective spin-wave mode has equal effective decay rates into the microwave and optical channels. In the low-loss limit $\Gamma_r\ll \Gamma_{\mu,o}$, the peak efficiency approaches unity, while the conversion bandwidth remains tunable through the effective Raman couplings.

\begin{figure}[t!]
\includegraphics[width=1.\columnwidth]{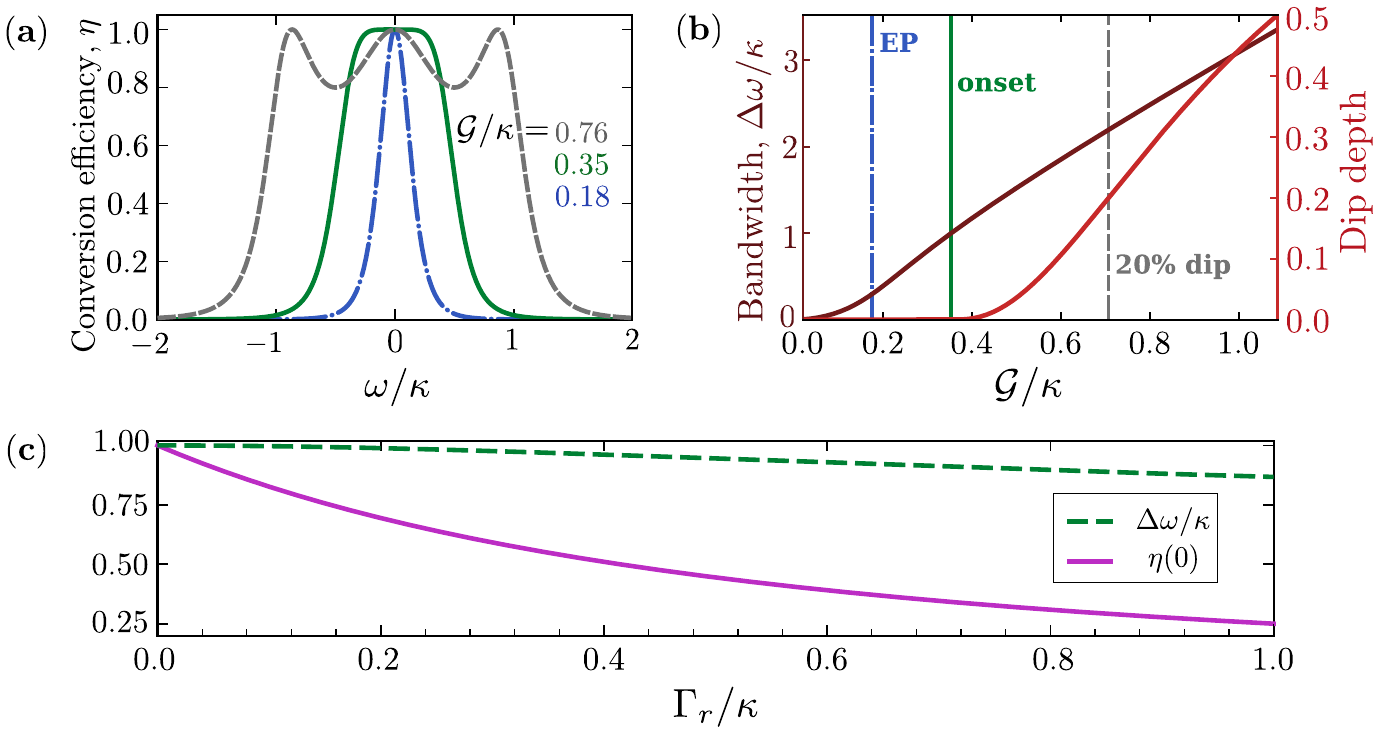}
\caption{Equal cavity-linewidth case, 
$\kappa_o=\kappa_\mu\equiv\kappa$.
(a) Conversion efficiency as a function of frequency offset from the cavity resonances for representative effective-coupling strengths $\mathcal G$, corresponding to the exceptional point (dash-dotted), the dip-onset point (solid), and a finite dip depth of $20\%$ (dashed). These three values are indicated in panel (b) using the same line styles. The spectra illustrate the transition from a single narrow resonance (dash-dotted) to a broadened plateau (solid) and a multi-peak structure (dashed) as $\mathcal G/\kappa$ increases. Under impedance-matched conditions with $\Gamma_r=0$, the peak efficiency remains unity at $\omega=0$, while increasing $\mathcal G$ redistributes spectral weight and generates off-center maxima separated by dips.
(b) Full-width-at-half-maximum bandwidth (brown) and dip depth (red), under impedance-matched conditions with $\Gamma_r=0$, as functions of $\mathcal G/\kappa$. The bandwidth increases monotonically with coupling, while spectral nonuniformity develops beyond the dip-onset point. Vertical lines, using the same line styles as in panel (a), mark the EP, the dip-onset point, and the point where the dip depth reaches $20\%$.
(c) On-resonance conversion efficiency (solid) and full-width-at-half-maximum bandwidth (dashed) as functions of the effective Rydberg-state linewidth, under impedance-matched conditions and evaluated at the dip-onset point, $\mathcal G/\kappa=1/(2\sqrt{2})$.}

\label{fig:symm}
\end{figure}

\textit{Symmetric case.---}%
We first consider the symmetric configuration $\kappa_\mu=\kappa_o\equiv\kappa$, for which the impedance-matching condition reduces to $\mathcal G_\mu=\mathcal G_o\equiv\mathcal G$. As shown in Fig.~\ref{fig:symm}(a), increasing $\mathcal G/\kappa$ drives the conversion spectrum from a narrow single resonance to a broadened, flat response before it eventually develops a three-peak structure, with two dips of equal depth separating the central peak from the side peaks. While the on-resonance efficiency remains fixed by the impedance-matching condition, increasing $\mathcal G/\kappa$ enhances the conversion bandwidth while simultaneously introducing spectral nonuniformity.

To quantify this behavior, Fig.~\ref{fig:symm}(b) shows the full-width-at-half-maximum bandwidth $\Delta\omega$ together with the depth of the dips between the central and side peaks. An exceptional point occurs at $\mathcal G/\kappa=1/(4\sqrt2)$, corresponding to the coalescence of the bright-sector poles discussed below, while the dip-onset point occurs at $\mathcal G/\kappa=1/(2\sqrt2)$, where the spectrum first develops off-center minima. At this onset, the bandwidth is simply $\Delta\omega=\kappa$. Beyond this point, the bandwidth continues to expand as the dips deepen. Once the dip depth reaches $50\%$, the above-half-maximum region becomes disconnected, rendering a single bandwidth ill-defined.

The spectral structure admits a simple interpretation in the bright--dark cavity basis. In the symmetric configuration, the dark mode $D=(a-b)/\sqrt2$ decouples from the collective spin-wave mode $R$, whereas the bright mode $B=(a+b)/\sqrt2$ hybridizes with it to form a pair of bright polaritons. The response is therefore governed by a single dark-mode pole, $\omega_D=-i\kappa/2$, alongside the bright polariton pair
\begin{equation}
\omega_\pm=
-\frac{i}{4}(\kappa+\Gamma_r)
\pm
\sqrt{2\mathcal G^2-\frac{(\kappa-\Gamma_r)^2}{16}}.
\end{equation}
The exceptional point occurs at the coalescence of the bright-polariton
poles. This is distinct from the onset of the dips, which arises from
interference between the dark-mode response and the split bright-polariton
responses.

%%%
%%%

A non-vanishing Rydberg decay further modifies the transduction performance. As shown in Fig.~\ref{fig:symm}(c), increasing $\Gamma_r$ reduces the on-resonance conversion efficiency according to Eq.~(\ref{eq:heta}) while also slightly narrowing the achievable conversion bandwidth. Nevertheless, for $\Gamma_r\ll\Gamma_{\mu,o}$, the on-resonance conversion efficiency remains close to unity. Within the decay model considered here, the Rydberg-decay channels contribute only vacuum noise, provided that they are effectively unoccupied, so the transducer remains quantum limited even for nonzero $\Gamma_r$ (see Supplemental Material for details).

\begin{figure}[t!]
\includegraphics[width=1.\columnwidth]{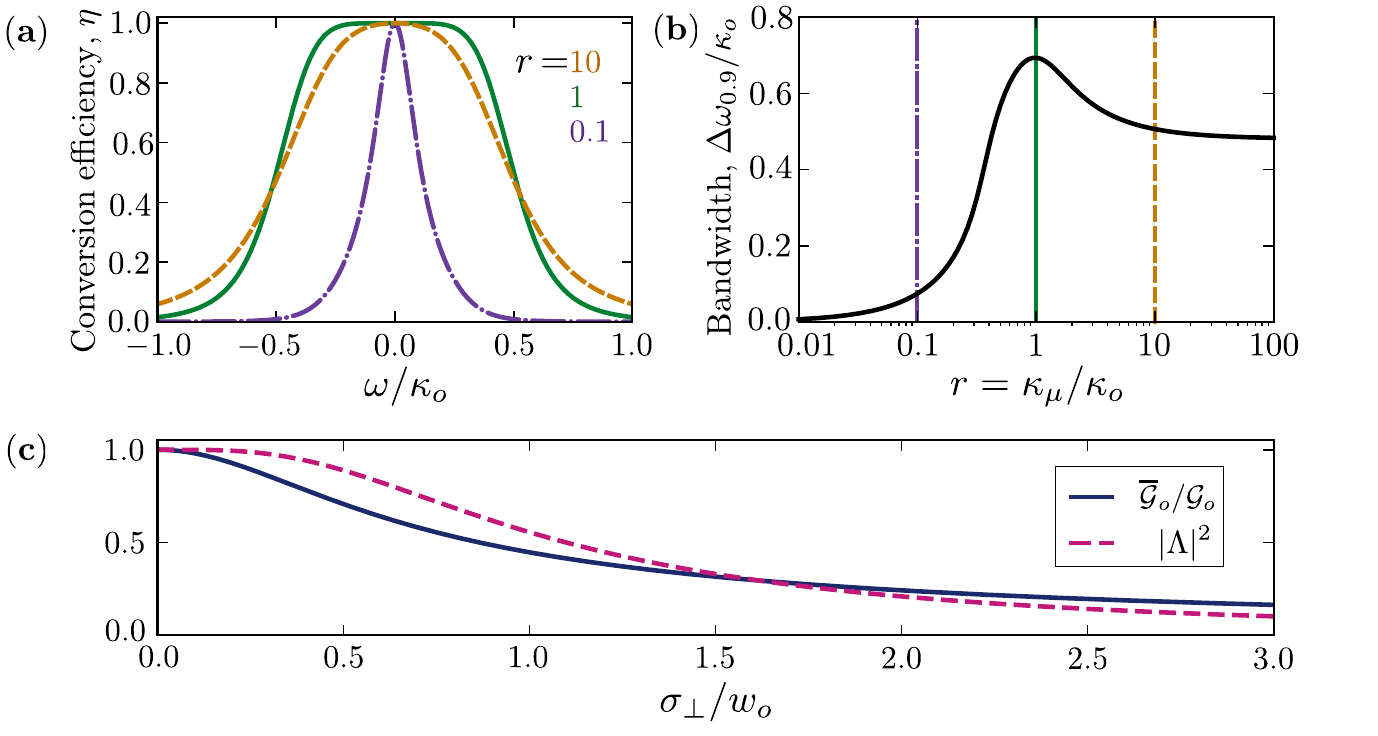}
\caption{Effects of cavity-linewidth asymmetry and spatially nonuniform coupling.
(a) Conversion efficiency as a function of frequency offset from the cavity
resonances for representative cavity-linewidth ratios
$r=0.1$ (dash-dotted), $1$ (solid), and $10$ (dashed), under
impedance-matched conditions with $\Gamma_r=0$ at $\mathcal G_o/\kappa_o=1/(2\sqrt{2})$. The symmetric case ($r=1$) exhibits a maximally flat broadband response, while asymmetric cases ($r\neq1$) lead to narrower spectral profiles.
(b) High-efficiency bandwidth $\Delta\omega_{0.9}$, defined by $\eta>0.9$, normalized to $\kappa_o$ as a function of the cavity-linewidth ratio $r$, using the same parameters as in panel (a). The bandwidth is maximized for matched cavity linewidths, $r=1$, and is generally set by the smaller of $\kappa_\mu$ and $\kappa_o$. Thus, for $r\ll1$, $\Delta\omega_{0.9}\propto\kappa_\mu$, giving $\Delta\omega_{0.9}/\kappa_o\propto r$, whereas for $r\gg1$, $\Delta\omega_{0.9}\propto\kappa_o$, and the normalized bandwidth saturates. Vertical lines mark the representative values $r=0.1,1,10$ shown in panel (a), using the same line styles.
(c) Optical collective-coupling reduction $\overline{\mathcal G}_o/\mathcal G_o$ and spin-wave-overlap factor $|\Lambda|^2$ for a Gaussian optical cavity mode as functions of the atomic-cloud size normalized to the optical-mode waist, $\sigma_\perp/w_o$. The microwave field is taken to be approximately uniform over the ensemble.}
\label{fig:asymm}
\end{figure}

\textit{Robustness to asymmetry and inhomogeneity.---}%
When the cavity-linewidth symmetry $\kappa_\mu=\kappa_o$ is lifted, the dark mode no longer fully decouples from the collective spin-wave mode, leading to a modified spectral structure. We characterize the asymmetry by the linewidth ratio $r\equiv\kappa_\mu/\kappa_o$. Away from the symmetric point $r=1$, impedance matching requires
unequal effective Raman couplings,
$\mathcal G_\mu/\mathcal G_o=\sqrt{r}$.
Figure~\ref{fig:asymm}(a) compares impedance-matched spectra for different values of $r$ at fixed $\mathcal G_o/\kappa_o=1/(2\sqrt{2})$, corresponding to the dip-onset coupling at $r=1$. Figure~\ref{fig:asymm}(b) shows the high-efficiency bandwidth $\Delta\omega_{0.9}$, defined by $\eta>0.9$, as a function of $r$. The distinct behavior in the $r\ll1$ and $r\gg1$ limits arises because the bandwidth is normalized to $\kappa_o$. For $r\ll1$, the narrower microwave cavity limits the input channel, causing out-of-band photons to be reflected and yielding
$\Delta\omega_{0.9}/\kappa_o\propto r$, or equivalently,
$\Delta\omega_{0.9}\propto\kappa_\mu$. For $r\gg1$, the optical cavity becomes the bottleneck; increasing $\kappa_\mu$ further cannot broaden the optical extraction window, so the normalized bandwidth saturates.

We finally consider spatially inhomogeneous atom--cavity coupling, which occurs when atoms experience different coupling amplitudes to the microwave and optical cavity modes. We write the position-dependent atom--cavity couplings as $g_{\alpha,l}=g_\alpha f_\alpha(\mathbf r_l)$, with $\alpha=\mu,o$, where $f_\alpha$ is normalized to unity at the field maximum. Spatial nonuniformity reduces the collective Raman couplings from their uniform-coupling values $\mathcal G_\alpha$ to root-sum-square values $\overline{\mathcal G}_\alpha$. Moreover, if the microwave and optical fields have different spatial profiles, they address partially distinct collective spin-wave modes.

For the trapped-atom cavity implementation considered in Fig.~\ref{fig:sketch}(b), the atomic cloud is positioned near the microwave coplanar-waveguide resonator and within the optical cavity mode. Provided that the spatial extent of the cloud is small compared with the characteristic length scale over which the microwave field varies, we can assume the microwave coupling to be approximately uniform, $f_\mu\simeq1$. This corresponds to trapping the atoms near a microwave-field maximum and at approximately the same distance from the chip surface. For the optical mode, we take the atomic cloud to be centered near the cavity-mode waist, with its longitudinal extent much smaller than the Rayleigh range, so that diffraction and Gouy-phase variations can be neglected. The remaining spatial variation is described by the transverse Gaussian profile
$f_o(\mathbf r_\perp)=e^{-|\mathbf r_\perp|^2/w_o^2}$, where $\mathbf r_\perp$ is the transverse position relative to the optical-mode axis and $w_o$ is the optical-mode waist. The normalized spin-wave-overlap factor $|\Lambda|^2 = \left|\langle f_\mu^*f_o\rangle\right|^2/\langle|f_\mu|^2\rangle\langle|f_o|^2\rangle $ quantifies the overlap between the collective spin waves addressed by the microwave and optical cavity modes. For imperfect overlap, the optical field also couples to an orthogonal collective spin-wave mode. The full two-spin-wave conversion spectrum for arbitrary $\Lambda$ is derived in the Supplemental Material.

Averaging the mode profiles over a Gaussian atomic density with transverse rms width $\sigma_\perp$ gives, under this uniform-microwave approximation, $\overline{\mathcal G}_\mu/\mathcal G_\mu\simeq1$, $\overline{\mathcal G}_o/\mathcal G_o=[1+4(\sigma_\perp/w_o)^2]^{-1/2}$, and a spin-wave-overlap factor $|\Lambda|^2=[1+4(\sigma_\perp/w_o)^2]/[1+2(\sigma_\perp/w_o)^2]^2$. Figure~\ref{fig:asymm}(c) plots the optical collective-Raman-coupling reduction $\overline{\mathcal G}_o/\mathcal G_o$ and the spin-wave-overlap factor $|\Lambda|^2$ as functions of $\sigma_\perp/w_o$, showing that tunable-bandwidth transduction remains robust when the atomic cloud is well contained within the optical waist. For $\sigma_\perp/w_o\simeq0.046$, corresponding to the transverse confinement reported in Ref.~\cite{Greve2022}, we obtain $\overline{\mathcal G}_o/\mathcal G_o\simeq0.996$ and $|\Lambda|^2\simeq0.99998$. For such a small spin-wave mismatch, the full two-spin-wave conversion spectrum remains close to the perfect-overlap result; representative spectra for several finite Rydberg-state linewidths are shown in the Supplemental Material. Recent atom-array platforms enable even more precise positioning of atoms near cavity-mode maxima~\cite{Liu2023,Seubert2025}.

\textit{Experimental considerations.---}%
The parameter regime considered here, including the near-symmetric cavity-linewidth regime discussed above, is compatible with existing optical-cavity, microwave-resonator, and Rydberg-atom platforms. Microwave and optical cavity linewidths can both be engineered to be in the MHz range. As a concrete microwave-cavity example, superconducting coplanar-waveguide resonators with widely tunable loaded quality factors can provide linewidths from the kHz scale to the tens-of-MHz range~\cite{Goppl2008}. Coherent coupling of Rydberg atoms to microwave fields in such resonators has also been demonstrated~\cite{Kaiser2022}. The MHz part of this range matches optical linewidths used in Rydberg-atom intracavity experiments~\cite{Sheng2017}. On the atomic side, microwave-dressed trapped Rydberg-state pairs suitable for the $\ket{s}\leftrightarrow\ket{r}$ transition have been identified for Rb atoms in Ref.~\cite{Tsiamis2026}. These states are long lived, with decay rates $\gamma_{r,s}/2\pi\lesssim1~\mathrm{kHz}$, so that the effective Rydberg-state linewidth can remain in the regime $\Gamma_r\ll\kappa_{\mu,o}$ under dispersive operation, making the scheme robust against atomic dissipation.

For a given experimental setup, the atom number, the bare atom--cavity
couplings, and the cavity linewidths are fixed device parameters. The
effective Raman couplings, however, scale as
$\mathcal G_o\propto\Omega_1/\Delta$ and
$\mathcal G_\mu\propto\Omega_2/\delta_s$, so their magnitudes and ratio
can be tuned \textit{in situ} through the driving-field Rabi frequencies
and detunings. For example, taking $N\simeq400$,
$\kappa_\mu/2\pi\simeq\kappa_o/2\pi\sim1~\mathrm{MHz}$,
$|g_o|/2\pi\simeq0.5~\mathrm{MHz}$~\cite{Greve2022},
which gives $\sqrt{N}|g_o|/2\pi\sim10~\mathrm{MHz}$,
$|g_\mu|/2\pi\sim10~\mathrm{MHz}$~\cite{Tsiamis2026,Kaiser2022},
and $|\Delta|/2\pi\simeq|\delta_s|/2\pi\sim100~\mathrm{MHz}$,
co-tuning $|\Omega_1|/2\pi$ from $0.1$ to $10~\mathrm{MHz}$ and
$|\Omega_2|/2\pi$ from $0.005$ to $0.5~\mathrm{MHz}$ preserves the
impedance-matching condition, $|\mathcal G_\mu|=|\mathcal G_o|$, while
broadening the conversion bandwidth from $0.8~\mathrm{kHz}$ to
approximately $3~\mathrm{MHz}$. This demonstrates that effective Raman couplings comparable to the cavity linewidths can be reached with an ensemble of $N\simeq400$ atoms while satisfying the adiabatic-elimination conditions throughout the tuning range (see Supplemental Material).

\textit{Conclusion.---}
We have proposed a cavity-based scheme for continuous-wave microwave-to-optical transduction mediated by a collective Rydberg excitation. In the impedance-matched regime, the system realizes a phase-preserving linear transducer that can maintain near-unit peak conversion efficiency while allowing the bandwidth to be tuned \textit{in situ} over more than three orders of magnitude, from spectrally selective sub-kilohertz operation to megahertz-scale broadband conversion.

The tunable-bandwidth mechanism admits a simple interpretation in terms of bright and dark cavity modes, where bandwidth enhancement arises from hybridization of the bright mode with the collective spin-wave mode and interference with the dark-mode response. We have shown that transducer performance remains robust throughout the tunable-bandwidth range under experimentally accessible levels of collective loss, cavity-linewidth asymmetry, and spatially inhomogeneous atom--cavity coupling. These results establish cavity-based Rydberg transduction as a promising platform for high-efficiency, quantum-limited, and reconfigurable microwave-to-optical interfaces in hybrid quantum networks.

\begin{acknowledgments}
\textit{Acknowledgments.---}We thank Jens Eisert for valuable discussions. We acknowledge financial support from Danmarks Grundforskningsfond (DNRF 139, Hy-Q Center for Hybrid Quantum Networks). I.T. acknowledges support from the Deutsche Forschungsgemeinschaft through the Emmy Noether program (Grant No. ME 4863/1-1). 
D.P. was supported by the HFRI project QuComPol (Grant No. 25442) and DFG FOR 5413 project QUSP (Grant No. 465199066). 
\end{acknowledgments}

\bibliography{references}

% ============================================================================
% Supplemental Material (appended for arXiv)
% ============================================================================
\clearpage
\onecolumngrid

% Supplemental figures use independent S-numbering.
\setcounter{figure}{0}
\renewcommand{\thefigure}{S\arabic{figure}}
\renewcommand{\theHfigure}{S\arabic{figure}}

\small

\begin{center}

{\Large\bfseries Supplemental Material}\\[6pt]

{\large\bfseries
Rydberg-mediated microwave-to-optical transduction with tunable bandwidth
}\\[8pt]

\normalsize
Iason Tsiamis\textsuperscript{1},
David Petrosyan\textsuperscript{2,3},
Anders S. S{\o}rensen\textsuperscript{4}\\[5pt]

\small\itshape

\parbox{0.64\textwidth}{\centering
\textsuperscript{1}Dahlem Center for Complex Quantum Systems and Fachbereich Physik, Freie Universität Berlin, 14195 Berlin, Germany
}\\[1pt]

\parbox{0.64\textwidth}{\centering
\textsuperscript{2}Institute of Electronic Structure and Laser and Center for Quantum
Science and Technologies, FORTH, 70013 Heraklion, Crete, Greece
}\\[1pt]

\parbox{0.65\textwidth}{\centering
\textsuperscript{3}Center for Quantum Science, Physikalisches Institut,
Eberhard Karls Universit\"at T\"ubingen, Auf der Morgenstelle 14,
72076 T\"ubingen, Germany
}\\[1pt]

\parbox{0.68\textwidth}{\centering
\textsuperscript{4}Center for Hybrid Quantum Networks (Hy-Q), The Niels Bohr Institute,
University of Copenhagen, DK-2100 Copenhagen \O, Denmark
}\\[1pt]

\end{center}

\vspace{1em}

\section*{S1. Hamiltonian and rotating frame}

We consider an ensemble of $N$ Rydberg atoms at positions $\mathbf r_l$ ($l=1,\dots,N$), each with four relevant internal states: the ground state $\ket{g}$, the excited state \(\ket{e}\), and two Rydberg states $\ket{s}$ and $\ket{r}$. A microwave cavity mode $a$ couples the transition $\ket{s}\leftrightarrow\ket{r}$ with single-atom coupling strength $g_\mu$, while an optical cavity mode $b$ couples $\ket{g}\leftrightarrow\ket{e}$ with coupling strength $g_o$. In addition, two classical control fields with Rabi frequencies $\Omega_1$ and $\Omega_2$ drive the transitions $\ket{r}\leftrightarrow\ket{e}$ and $\ket{s}\leftrightarrow\ket{g}$, respectively. We neglect interatomic interactions involving Rydberg excitations in the states \(\ket{s}\) and \(\ket{r}\), assuming sufficiently weak input fields and/or sufficiently low atomic densities that interaction-induced level shifts are negligible.

The corresponding time-dependent system Hamiltonian reads
\begin{align}
\frac{H}{\hbar}
&=
\omega_\mu a^\dagger a
+\omega_o b^\dagger b
+\sum_l\left(
\omega_s \sigma_{ss}^{(l)}
+\omega_r \sigma_{rr}^{(l)}
+\omega_e \sigma_{ee}^{(l)}
\right)
\nonumber\\
&\quad
+\sum_l \Big[
g_\mu a^\dagger \sigma_{sr}^{(l)}
+g_o b^\dagger e^{-i\mathbf k_o\cdot \mathbf r_l}\sigma_{ge}^{(l)}
+\Omega_1 e^{-i\omega_1 t} e^{i\mathbf k_1\cdot \mathbf r_l}\sigma_{re}^{(l)}
+\Omega_2 e^{-i\omega_2 t} e^{i\mathbf k_2\cdot \mathbf r_l}\sigma_{sg}^{(l)}
+\mathrm{H.c.}
\Big],
\tag{S1}
\end{align}
where $\sigma_{\alpha\beta}^{(l)}=\ket{\alpha}_l\!\bra{\beta}$, $\omega_\mu$ and $\omega_o$ are the microwave and optical cavity mode frequencies, and $\omega_1$ and $\omega_2$ are the frequencies of the two classical drives. Assuming spatially uniform couplings, we take $g_\mu$, $g_o$, $\Omega_1$, and $\Omega_2$ to be real by absorbing their constant phases into the operator
definitions.

For a traveling-wave optical cavity mode, the atom--cavity interaction takes the form given in Eq.~(S1). For a standing-wave optical cavity, the coupling
instead contains the position-dependent standing-wave mode function. If the atomic ensemble is tightly confined near a single antinode, this mode function is
approximately constant over the atoms, and the same uniform-coupling model is recovered with $g_o$ replaced by its local value and the traveling-wave
phase factor $e^{-i\mathbf{k}_o\cdot\mathbf r_l}$ omitted. For an ensemble spanning multiple antinodes, the full standing-wave mode function, including any oscillating signs, must be retained and treated as a spatially nonuniform atom--cavity coupling.

To remove the explicit time dependence of Eq.~(S1), we move to a rotating frame defined by the unitary transformation
\begin{align}
U(t)=\exp\!\left\{-
it\left[
\omega_\mu a^\dagger a
+\omega_o b^\dagger b
+\sum_l\Big(
\omega_2 \sigma_{ss}^{(l)}
+(\omega_o+\omega_1)\sigma_{rr}^{(l)}
+\omega_o \sigma_{ee}^{(l)}
\Big)
\right]
\right\},
\tag{S2}
\end{align}
under which the transformed Hamiltonian $\tilde H = U^\dagger H U - i\hbar\, U^\dagger \dot U$ becomes
\begin{align}
\frac{\tilde H}{\hbar}
&=
\sum_l\left(
\delta_s \sigma_{ss}^{(l)}
+\delta_r \sigma_{rr}^{(l)}
+\Delta \sigma_{ee}^{(l)}
\right)
\nonumber\\
&\quad
+\sum_l \Big[
g_\mu a^\dagger \sigma_{sr}^{(l)} e^{i\delta t}
+g_o b^\dagger e^{-i\mathbf k_o\cdot \mathbf r_l}\sigma_{ge}^{(l)}
+\Omega_1 e^{i\mathbf k_1\cdot \mathbf r_l}\sigma_{re}^{(l)}
+\Omega_2 e^{i\mathbf k_2\cdot \mathbf r_l}\sigma_{sg}^{(l)}
+\mathrm{H.c.}
\Big],
\tag{S3}
\end{align}
with the corresponding detunings defined as
\begin{align}
\delta_s &= \omega_s-\omega_2,
&
\delta_r &= \omega_r-(\omega_o+\omega_1),
\nonumber\\
\Delta &= \omega_e-\omega_o,
&
\delta &= \omega_\mu+\omega_2-\omega_1-\omega_o.
\tag{S4}
\end{align}

Under the four-photon resonance condition $\delta=0$, the Hamiltonian becomes time independent. We can then absorb the spatial phases into the local atomic operators via a gauge transformation, leaving only the net phase-mismatch wavevector
\begin{align}
\mathbf K = \mathbf k_o + \mathbf k_1 - \mathbf k_2.
\tag{S5}
\end{align}
The rotating-frame Hamiltonian then takes the form
\begin{align}
\frac{\tilde H}{\hbar}
&=
\sum_l\left(
\delta_s \sigma_{ss}^{(l)}
+\delta_r \sigma_{rr}^{(l)}
+\Delta \sigma_{ee}^{(l)}
\right)
\nonumber\\
&\quad
+\sum_l \Big[
g_\mu a^\dagger \sigma_{sr}^{(l)}
+g_o b^\dagger e^{-i\mathbf K\cdot \mathbf r_l}\sigma_{ge}^{(l)}
+\Omega_1 \sigma_{re}^{(l)}
+\Omega_2 \sigma_{sg}^{(l)}
+\mathrm{H.c.}
\Big],
\tag{S6}
\end{align}
which matches Eq.~(1) of the main text.

\section*{S2. Adiabatic elimination}

We consider the dispersive regime in which the states $\ket s$ and $\ket e$ remain only virtually populated due to the large detunings $|\delta_s|$ and $|\Delta|$. Their dynamics therefore evolve on much faster timescales than those associated with $\{\ket g,\ket r\}$, allowing the far-detuned states to be adiabatically eliminated. The system then reduces to an effective description within the slow manifold spanned by $\{\ket g,\ket r\}$, obtained using the effective-operator formalism of Ref.~\cite{Reiter2012}. 

The dynamics of the full atom--cavity system are described by the master equation
\begin{equation}
\dot\rho
=
-\frac{i}{\hbar}[\tilde H,\rho]
+\sum_k \mathcal D[L_k]\rho,
\tag{S7}
\end{equation}
with Lindblad dissipator
\begin{equation}
\mathcal D[L]\rho
=
L\rho L^\dagger
-\frac{1}{2}L^\dagger L\rho
-\frac{1}{2}\rho L^\dagger L .
\tag{S8}
\end{equation}
The dissipative channels are described by the jump operators
\begin{align}
L_{s,l} &= \sqrt{\gamma_s}\,\sigma_{gs}^{(l)},
&
L_{r,l} &= \sqrt{\gamma_r}\,\sigma_{gr}^{(l)},
&
L_{e,l} &= \sqrt{\gamma_e}\,\sigma_{ge}^{(l)},
\tag{S9}
\\
L_\mu &= \sqrt{\kappa_\mu}\,a,
&
L_o &= \sqrt{\kappa_o}\,b ,
\tag{S10}
\end{align}
where $\gamma_s$, $\gamma_r$, and $\gamma_e$ are the decay rates of the atomic states $\ket{s}$, $\ket{r}$, and $\ket{e}$ into the ground state $\ket{g}$, while $\kappa_\mu$ and $\kappa_o$ denote the linewidths of the microwave and optical cavity modes, respectively.

The elimination is valid provided the couplings to the far-detuned states remain small compared with the corresponding detunings,
\begin{equation}
|\delta_s|
\gg
|\Omega_2|\sqrt{N},
\quad
|\delta_s|
\gg
|g_\mu| \sqrt{\langle a^\dagger a + 1\rangle},
\qquad
|\Delta|
\gg
|\Omega_1|,
\quad
|\Delta|
\gg
|g_o| \sqrt{N\,\langle b^\dagger b + 1\rangle}.
\tag{S11}
\end{equation}
The factors of $\sqrt{N}$ reflect collective enhancement in the symmetric subspace. Throughout, we work in the weak-excitation regime, where the number of atomic excitations remains much smaller than $N$ and almost all atoms remain in the ground state $\ket g$.

We decompose the rotating-frame Hamiltonian as
\begin{equation}
\tilde H = H_g + H_e + V_+ + V_-,
\tag{S12}
\end{equation}
where
\begin{equation}
\frac{H_g}{\hbar}
=
\delta_r \sum_l \sigma_{rr}^{(l)},
\tag{S13}
\end{equation}
\begin{equation}
\frac{H_e}{\hbar}
=
\delta_s \sum_l \sigma_{ss}^{(l)}
+
\Delta \sum_l \sigma_{ee}^{(l)},
\tag{S14}
\end{equation}
and
\begin{align}
\frac{V_+}{\hbar}
&=
\sum_l \Big[
\Omega_2 \sigma_{sg}^{(l)}
+g_\mu a^\dagger \sigma_{sr}^{(l)}
+g_o b\, e^{i\mathbf K\cdot\mathbf r_l}\sigma_{eg}^{(l)}
+\Omega_1 \sigma_{er}^{(l)}
\Big],
\tag{S15}
\\
V_- &= V_+^\dagger .
\tag{S16}
\end{align}
Here $V_+$ transfers population from the slow manifold to the eliminated manifold, while $V_-$ returns it.

Within the effective-operator formalism, the non-Hermitian Hamiltonian governing the virtual dynamics in the eliminated subspace is
\begin{equation}
H_{\rm NH}
=
H_e
-\frac{i\hbar}{2}\sum_l
\left(
L_{s,l}^\dagger L_{s,l}
+
L_{e,l}^\dagger L_{e,l}
\right),
\tag{S17}
\end{equation}
which evaluates to
\begin{equation}
\frac{H_{\rm NH}}{\hbar}
=
\left(\delta_s-\frac{i\gamma_s}{2}\right)\sum_l \sigma_{ss}^{(l)}
+
\left(\Delta-\frac{i\gamma_e}{2}\right)\sum_l \sigma_{ee}^{(l)} .
\tag{S18}
\end{equation}
We define the complex detunings
\begin{equation}
\tilde\delta_s \equiv \delta_s-\frac{i\gamma_s}{2},
\qquad
\tilde\Delta \equiv \Delta-\frac{i\gamma_e}{2}.
\tag{S19}
\end{equation}
Since the $\ket s$ and $\ket e$ states do not couple directly, $H_{\rm NH}^{-1}$ is diagonal,
\begin{equation}
H_{\rm NH}^{-1}
=
\frac{1}{\hbar}
\left(
\frac{\sum_l \sigma_{ss}^{(l)}}{\tilde\delta_s}
+
\frac{\sum_l \sigma_{ee}^{(l)}}{\tilde\Delta}
\right).
\tag{S20}
\end{equation}

Adiabatic elimination yields the effective Hamiltonian and jump operators acting within the slow manifold:
\begin{align}
H_{\rm eff}
&=
H_g
-\frac{1}{2}
V_-
\left(
H_{\rm NH}^{-1}
+
H_{\rm NH}^{-1\dagger}
\right)
V_+,
\tag{S21}
\\
L_{\rm eff}^{(k)}
&=
L_k H_{\rm NH}^{-1}V_+,
\tag{S22}
\end{align}
where $k$ runs over the eliminated-subspace decay channels $L_{s,l}$ and $L_{e,l}$. Evaluating Eq.~(S22) yields
\begin{align}
L_{{\rm eff},s,l}
&=
\sqrt{\gamma_s}\left(
\frac{\Omega_2}{\tilde\delta_s}\sigma_{gg}^{(l)}
+
\frac{g_\mu}{\tilde\delta_s}a^\dagger \sigma_{gr}^{(l)}
\right),
\tag{S23}
\\
L_{{\rm eff},e,l}
&=
\sqrt{\gamma_e}\left(
\frac{g_o}{\tilde\Delta} b\,e^{i\mathbf K\cdot\mathbf r_l}\sigma_{gg}^{(l)}
+
\frac{\Omega_1}{\tilde\Delta}\sigma_{gr}^{(l)}
\right).
\tag{S24}
\end{align}

The terms proportional to $\sigma_{gr}^{(l)}$ describe induced decay of the effective Rydberg excitation arising from virtual population of the eliminated states. The corresponding induced linewidth scales as
\begin{equation}
\Gamma_r^{\rm(ind)}
=
\frac{\gamma_s g_\mu^2}{|\tilde\delta_s|^2}\,\langle a a^\dagger\rangle
+
\frac{\gamma_e \Omega_1^2}{|\tilde\Delta|^2}.
\tag{S25}
\end{equation}
The total Rydberg-state linewidth is therefore
\begin{equation}
\Gamma_r \equiv \gamma_r + \Gamma_r^{\rm(ind)},
\tag{S26}
\end{equation}
where $\gamma_r$ is the bare decay rate of $\ket r$.

In the weak-excitation regime, $\sigma_{gg}^{(l)}\simeq1$, so that the term
proportional to $\sigma_{gg}^{(l)}$ in Eq.~(S23) is approximately
proportional to the identity, and its standalone contribution to the
Lindblad dissipator vanishes. It cannot, however, be discarded directly
from the effective jump operator, since cross terms with the second
contribution can generate a coherent correction. As shown below, these
terms are parametrically suppressed in the far-detuned regime. By contrast,
the term proportional to $b\sigma_{gg}^{(l)}$ in Eq.~(S24) produces an
additional atom-induced optical-cavity loss rate
$N\gamma_e|g_o|^2/|\tilde{\Delta}|^2$. Equation~(S11) ensures that this
loss is parametrically suppressed, since
$N|g_o|^2/|\tilde{\Delta}|^2\ll1$, and it can be neglected under the additional condition
$N\gamma_e|g_o|^2/|\tilde{\Delta}|^2\ll\kappa_o$.

Evaluating Eq.~(S21), the coherent part of the effective dynamics becomes
\begin{align}
\frac{H_{\rm eff}}{\hbar}
&=
\delta_{\rm eff}\sum_l \sigma_{rr}^{(l)}
+\sum_l \Big[
G_\mu^{(1)}
a^\dagger \sigma_{gr}^{(l)} 
+
G_o^{(1)}
b^\dagger e^{-i\mathbf K\cdot \mathbf r_l}\sigma_{gr}^{(l)}
+\mathrm{H.c.}\Big],
\tag{S27}
\end{align}
with single-atom effective Raman couplings
\begin{equation}
G_\mu^{(1)}
=
-\frac{g_\mu \Omega_2}{\tilde\delta_s},
\qquad
G_o^{(1)}
=
-\frac{g_o \Omega_1}{\tilde\Delta},
\tag{S28}
\end{equation}
and the effective detuning
\begin{equation}
\delta_{\rm eff}
=
\delta_r
-
\frac{\Omega_1^2\,\Delta}{\Delta^2+(\gamma_e/2)^2}
+
\frac{\Omega_2^2\,\delta_s}{\delta_s^2+(\gamma_s/2)^2}
+
\frac{g_o^2\,\Delta}{\Delta^2+(\gamma_e/2)^2}\langle b^\dagger b\rangle
-
\frac{g_\mu^2\,\delta_s}{\delta_s^2+(\gamma_s/2)^2}\langle a a^\dagger\rangle.
\tag{S29}
\end{equation}
The terms proportional to $\Omega_1^2$ and $\Omega_2^2$ correspond to AC Stark shifts induced by the classical drives, while the terms proportional to $g_o^2$ and $g_\mu^2$ describe cavity-induced dispersive shifts.

In the far-detuned regime, $|\delta_s|\gg\gamma_s$ and $|\Delta|\gg\gamma_e$, we can substitute $\tilde\Delta\approx\Delta$ and $\tilde\delta_s\approx\delta_s$ in Eqs.~(S25), (S28), (S29). The effective detuning then reduces to
\begin{equation}
\delta_{\rm eff}
\approx
\delta_r
-\frac{\Omega_1^2}{\Delta}
+\frac{\Omega_2^2}{\delta_s}
+\frac{g_o^2}{\Delta}\,\langle b^\dagger b\rangle
-\frac{g_\mu^2}{\delta_s}\,\langle a a^\dagger\rangle.
\tag{S30}
\end{equation}
Similarly, the single-atom effective Raman couplings become
\begin{equation}
G_\mu^{(1)}
\approx
-\frac{g_\mu \Omega_2}{\delta_s},
\qquad
G_o^{(1)}
\approx
-\frac{g_o \Omega_1}{\Delta},
\tag{S31}
\end{equation}
while the effective Rydberg-state linewidth reduces to
\begin{equation}
\Gamma_r
\approx
\gamma_r
+
\frac{\gamma_e \Omega_1^2}{\Delta^2}+
\frac{\gamma_s g_\mu^2}{\delta_s^2}\,\langle a a^\dagger\rangle.
\tag{S32}
\end{equation}

In the weak-excitation regime, almost all atoms remain in the ground state such that
$\langle \sigma_{gg}^{(l)} \rangle \approx 1$.
To capture the collective coupling of the atomic ensemble to the Raman driving fields, we introduce the collective Rydberg spin-wave operator
\begin{equation}
R \equiv \frac{1}{\sqrt N}\sum_l e^{-i\mathbf K\cdot\mathbf r_l}\sigma^{(l)}_{gr}.
\tag{S33}
\end{equation}
Within this regime, this collective operator approximately satisfies the bosonic commutation relation $[R,R^\dagger]\approx1$. The effective Hamiltonian then becomes
\begin{align}
\frac{H_{\rm eff}}{\hbar}
&=
\delta_{\rm eff} \sum_l\sigma_{rr}^{(l)}
+
\left(
G_\mu^{(1)}
a^\dagger \sum_l \sigma_{gr}^{(l)}
+
\sqrt N\,G_o^{(1)}
b^\dagger R
+\mathrm{H.c.}
\right).
\tag{S34}
\end{align}

Under near-phase-matching conditions $\mathbf K\approx0$, the coherent dynamics further reduce to the linear three-mode effective Hamiltonian:
\begin{equation}
\frac{H_{\mathrm{eff}}}{\hbar}
=
\delta_{\rm eff} R^\dagger R
+
\left(
\mathcal G_\mu a^\dagger R
+
\mathcal G_o b^\dagger R
+ \mathrm{H.c.}
\right),
\tag{S35}
\end{equation}
with collectively enhanced Raman couplings
\begin{equation}
\mathcal G_\mu=\sqrt N\,G_\mu^{(1)},
\qquad
\mathcal G_o=\sqrt N\,G_o^{(1)}.
\tag{S36}
\end{equation}

The remaining cross terms between the two contributions in
Eqs.~(S23) and~(S24) are also neglected in the far-detuned regime. For
Eq.~(S23), the cross term involving the
$\Omega_2\sigma_{gg}^{(l)}$ contribution is equivalent to a Hamiltonian
correction to the microwave Raman coupling, with characteristic strength
$\sqrt{N}\gamma_s g_\mu\Omega_2/(2|\tilde\delta_s|^2)$ in the amplitude
equations. This is smaller than the coherent microwave Raman coupling by
$\gamma_s/(2|\delta_s|)\ll1$. For Eq.~(S24), the cross term instead produces
a correlated dissipative coupling between the optical cavity mode $b$ and the
collective Rydberg mode $R$, with characteristic strength
$\sqrt{N}\gamma_e g_o\Omega_1/(2|\tilde\Delta|^2)$ in the amplitude equations.
This is smaller than the coherent optical Raman coupling by
$\gamma_e/(2|\Delta|)\ll1$.

After neglecting these cross terms together with the small atom-induced
optical-cavity loss discussed above, the reduced dissipative dynamics are
described by the jump operators
\begin{equation}
L_\mu=\sqrt{\kappa_\mu}\,a,
\qquad
L_o=\sqrt{\kappa_o}\,b,
\qquad
L_R=\sqrt{\Gamma_r}\,R .
\tag{S37}
\end{equation}

\section*{S3. Input--output theory and added noise}

The reduced collective description obtained in Sec.~S2 provides the starting point for the input--output analysis of the microwave-to-optical conversion spectrum. Using the effective Hamiltonian~(S35) and jump operators~(S37), the Heisenberg--Langevin equations follow from
\begin{equation}
\dot O
=
\frac{i}{\hbar}[H_{\rm eff},O]
+\sum_j
\left(
L_j^\dagger O L_j
-\frac12\{L_j^\dagger L_j,O\}
\right)
+\hat F_O ,
\tag{S38}
\end{equation}
where $\hat F_O$ denotes the corresponding input noise operator. Working in the rotating frame of the microwave and optical cavity modes then gives
\begin{align}
\dot a
&=
-\frac{\kappa_\mu}{2}a
-i\mathcal G_\mu R
+\sqrt{\kappa_\mu}\, a_{\rm in},
\tag{S39}
\\
\dot b
&=
-\frac{\kappa_o}{2}b
-i\mathcal G_o R
+\sqrt{\kappa_o}\, b_{\rm in},
\tag{S40}
\\
\dot R
&=
-\left(\frac{\Gamma_r}{2}+i\delta_{\rm eff}\right)R
-i\mathcal G_\mu a
-i\mathcal G_o b
+\sqrt{\Gamma_r}\, R_{\rm in}.
\tag{S41}
\end{align}
Here $a_{\rm in}$ and $b_{\rm in}$ are the input fields associated with the microwave and optical cavities, respectively, while $R_{\rm in}$ describes the noise operator associated with the collective Rydberg decay channel. The cavity output fields are given by the standard input--output relations of Ref.~\cite{Gardiner1985}
\begin{equation}
a_{\rm out}=a_{\rm in}-\sqrt{\kappa_\mu}\,a,
\qquad
b_{\rm out}=b_{\rm in}-\sqrt{\kappa_o}\,b.
\tag{S42}
\end{equation}

Using the Fourier convention
\begin{equation}
O(\omega)=\int_{-\infty}^{\infty} dt\, e^{i\omega t} O(t),
\tag{S43}
\end{equation}
where $\omega$ denotes the frequency offset from the cavity resonances, the Heisenberg--Langevin equations become
\begin{equation}
M(\omega)
\begin{pmatrix}
a(\omega)\\
b(\omega)\\
R(\omega)
\end{pmatrix}
=
i
\begin{pmatrix}
\sqrt{\kappa_\mu}\, a_{\rm in}(\omega)\\
\sqrt{\kappa_o}\, b_{\rm in}(\omega)\\
\sqrt{\Gamma_r}\, R_{\rm in}(\omega)
\end{pmatrix},
\tag{S44}
\end{equation}
with
\begin{equation}
M(\omega)=
\begin{pmatrix}
\omega+i\kappa_\mu/2 & 0 & -\mathcal G_\mu\\
0 & \omega+i\kappa_o/2 & -\mathcal G_o\\
-\mathcal G_\mu & -\mathcal G_o & \omega-\delta_{\rm eff}+i\Gamma_r/2
\end{pmatrix}.
\tag{S45}
\end{equation}
The corresponding determinant is
\begin{equation}
\mathcal D(\omega)\equiv \det M(\omega)
=
\left(\omega+\frac{i\kappa_\mu}{2}\right)
\left(\omega+\frac{i\kappa_o}{2}\right)
\left(\omega-\delta_{\rm eff}+\frac{i\Gamma_r}{2}\right)
-\left(\omega+\frac{i\kappa_\mu}{2}\right)\mathcal G_o^2
-\left(\omega+\frac{i\kappa_o}{2}\right)\mathcal G_\mu^2 .
\tag{S46}
\end{equation}

Solving Eq.~(S44) and using $b_{\rm out}=b_{\rm in}-\sqrt{\kappa_o}\,b$, the coefficient of $a_{\rm in}(\omega)$ in $b_{\rm out}(\omega)$ defines the microwave-to-optical conversion amplitude,
\begin{equation}
S_{o\mu}(\omega)
=
-i\sqrt{\kappa_\mu\kappa_o}\,
\frac{\mathcal G_\mu\mathcal G_o}{\mathcal D(\omega)} .
\tag{S47}
\end{equation}
The conversion efficiency is
\begin{equation}
\eta(\omega)=|S_{o\mu}(\omega)|^2.
\tag{S48}
\end{equation}

The denominator $\mathcal D(\omega)$ controls the spectral shape, whereas the peak conversion is fixed by how strongly the microwave and optical channels load the collective spin mode. On resonance, $\omega=\delta_{\rm eff}=0$, one finds
\begin{equation}
S_{o\mu}(0)
=
\frac{2\sqrt{\Gamma_\mu\Gamma_o}}
{\Gamma_r+\Gamma_\mu+\Gamma_o},
\tag{S49}
\end{equation}
where
\begin{equation}
\Gamma_\mu \equiv \frac{4\mathcal G_\mu^2}{\kappa_\mu},
\qquad
\Gamma_o \equiv \frac{4\mathcal G_o^2}{\kappa_o}
\tag{S50}
\end{equation}
are the effective microwave and optical loading rates of the collective spin-wave mode. Accordingly,
\begin{equation}
\eta(0)
=
\frac{4\Gamma_\mu\Gamma_o}
{(\Gamma_r+\Gamma_\mu+\Gamma_o)^2}.
\tag{S51}
\end{equation}
This is maximized when the two cavity channels load the collective spin equally,
\begin{equation}
\Gamma_\mu=\Gamma_o
\qquad\Longleftrightarrow\qquad
\frac{\mathcal G_\mu^2}{\kappa_\mu}
=
\frac{\mathcal G_o^2}{\kappa_o}.
\tag{S52}
\end{equation}
Equation~(S52) is the impedance-matching condition for the passive converter. It fixes the optimal loading of the spin mode, while the pole structure determines the bandwidth and lineshape of the conversion spectrum. Finite $\Gamma_r$ lowers the peak conversion according to Eq.~(S51) and reshapes the spectrum through the frequency dependence of $\mathcal D(\omega)$.

The full optical output field may be written as
\begin{equation}
b_{\rm out}(\omega)
=
S_{o\mu}(\omega)a_{\rm in}(\omega)
+
S_{oo}(\omega)b_{\rm in}(\omega)
+
S_{or}(\omega)R_{\rm in}(\omega).
\tag{S53}
\end{equation}
We consider microwave-to-optical conversion with the microwave port driven by the input signal, while the optical input port and the Rydberg loss channel contribute only noise fields. For uncorrelated baths, the normally ordered added noise referred to the microwave input signal is
\begin{equation}
n_{\rm add}(\omega)
=
\frac{
|S_{oo}(\omega)|^2 n_o
+
|S_{or}(\omega)|^2 n_r
}
{|S_{o\mu}(\omega)|^2},
\tag{S54}
\end{equation}
where $n_o=\langle b_{\rm in}^\dagger b_{\rm in}\rangle$ and
$n_r=\langle R_{\rm in}^\dagger R_{\rm in}\rangle$ denote the thermal occupations of the optical input and Rydberg loss channels, respectively. Thus, for vacuum optical input and an unoccupied Rydberg loss channel, $n_o=n_r=0$, so the added noise vanishes, $n_{\rm add}=0$. Reaching $n_r\simeq0$ for the relevant low-energy Rydberg
transitions requires cryogenic operation, consistent with the temperatures
already needed to suppress thermal microwave noise. Under these conditions,
finite Rydberg loss, $\Gamma_r\neq0$, suppresses the conversion efficiency
without contributing added noise.

\section*{S4. Bright--dark basis}

The microwave-to-optical conversion spectrum derived in Sec.~S3 admits a simple interpretation in terms of bright and dark cavity superposition modes. We introduce these modes through the definitions
\begin{equation}
\mathcal G_B \equiv \sqrt{\mathcal G_\mu^2+\mathcal G_o^2},
\qquad
B \equiv \frac{\mathcal G_\mu a+\mathcal G_o b}{\mathcal G_B},
\qquad
D \equiv \frac{\mathcal G_o a-\mathcal G_\mu b}{\mathcal G_B}.
\tag{S55}
\end{equation}
Here $B$ is the cavity superposition that couples directly to the collective spin-wave mode, while $D$ is orthogonal to it.

At the level of the coherent effective Hamiltonian~(S35), the collective spin--cavity interaction
\begin{equation}
\frac{H_{\rm int}}{\hbar}
=
\mathcal G_\mu \bigl(a^\dagger R+aR^\dagger\bigr)
+
\mathcal G_o \bigl(b^\dagger R+bR^\dagger\bigr)
\tag{S56}
\end{equation}
can be written in terms of the bright and dark cavity modes as
\begin{equation}
\frac{H_{\rm int}}{\hbar}
=
\mathcal G_B\bigl(B^\dagger R+BR^\dagger\bigr),
\tag{S57}
\end{equation}
so that only the bright mode couples directly to the collective spin-wave mode. In the open-system response, however, a linewidth mismatch between the microwave and optical cavities can still induce mixing between $B$ and $D$, as captured by the non-Hermitian parameter $\zeta$ below.

In the $(B,D,R)$ basis, the non-Hermitian response kernel corresponding to Eq.~(S45) takes the form
\begin{equation}
M_{BDR}(\omega)=
\begin{pmatrix}
\omega-z_B & -\zeta & -\mathcal G_B\\
-\zeta & \omega-z_D & 0\\
-\mathcal G_B & 0 & \omega-z_r
\end{pmatrix},
\tag{S58}
\end{equation}
where
\begin{equation}
z_B
=
\frac{\mathcal G_\mu^2 z_\mu+\mathcal G_o^2 z_o}{\mathcal G_B^2},
\qquad
z_D
=
\frac{\mathcal G_o^2 z_\mu+\mathcal G_\mu^2 z_o}{\mathcal G_B^2},
\tag{S59}
\end{equation}
and
\begin{equation}
\zeta
=
\frac{\mathcal G_\mu\mathcal G_o}{\mathcal G_B^2}(z_\mu-z_o).
\tag{S60}
\end{equation}
Here $z_\mu=-i\kappa_\mu/2$, $z_o=-i\kappa_o/2$, and $z_r=\delta_{\rm eff}-i\Gamma_r/2$ denote the complex eigenfrequencies of the microwave, optical, and collective Rydberg modes in the rotating frame of the cavity modes. The parameter $\zeta$ describes bright--dark mixing and is controlled by the mismatch between the complex cavity frequencies.

The determinant appearing in Eq.~(S46), which controls the spectral structure of the conversion, can then be written as
\begin{equation}
\mathcal D(\omega)
=
(\omega-z_D)\Big[(\omega-z_B)(\omega-z_r)-\mathcal G_B^2\Big]
-\zeta^2(\omega-z_r).
\tag{S61}
\end{equation}
When the linewidths of the two cavity modes are matched,
\begin{equation}
\kappa_\mu=\kappa_o\equiv\kappa,
\tag{S62}
\end{equation}
(equivalently $z_\mu=z_o$), one has $\zeta=0$ and the determinant factorizes as
\begin{equation}
\mathcal D(\omega)
=
(\omega-z_D)\Big[(\omega-z_B)(\omega-z_r)-\mathcal G_B^2\Big].
\tag{S63}
\end{equation}
The spectral response is then governed by one dark pole,
\begin{equation}
\omega_D=z_D,
\tag{S64}
\end{equation}
and a bright polariton pair,
\begin{equation}
\omega_\pm
=
\frac{z_B+z_r}{2}
\pm
\sqrt{
\mathcal G_B^2+\frac{(z_B-z_r)^2}{4}
}.
\tag{S65}
\end{equation}

Under impedance matching [Eq.~(S52)], this further implies
\begin{equation}
\mathcal G_\mu=\mathcal G_o\equiv\mathcal G,
\tag{S66}
\end{equation}
and the bright and dark modes reduce to the symmetric and antisymmetric cavity superpositions
\begin{equation}
B=\frac{a+b}{\sqrt2},
\qquad
D=\frac{a-b}{\sqrt2}.
\tag{S67}
\end{equation}
The dark pole is
\begin{equation}
\omega_D=-\frac{i\kappa}{2},
\tag{S68}
\end{equation}
while the bright polariton poles become
\begin{equation}
\omega_\pm
=
-\frac{i}{4}(\kappa+\Gamma_r)
\pm
\sqrt{
2\mathcal G^2-\frac{(\kappa-\Gamma_r)^2}{16}
}.
\tag{S69}
\end{equation}

This decomposition explains why the exceptional point of the bright polariton poles and the onset of spectral dips in the observable conversion spectrum do not coincide. The exceptional point occurs when the square root in Eq.~(S69) vanishes. In the low-loss limit $\Gamma_r\ll\kappa$, this gives
\begin{equation}
\frac{\mathcal G}{\kappa}
=
\frac{1}{4\sqrt2}.
\tag{S70}
\end{equation}
By contrast, the observable conversion spectrum depends on the coherent interference between the dark pole and the bright polariton pair. The spectrum therefore remains single-peaked beyond the exceptional point and develops off-center minima only at the larger coupling
\begin{equation}
\frac{\mathcal G}{\kappa}
=
\frac{1}{2\sqrt2}.
\tag{S71}
\end{equation}
For larger Raman coupling $\mathcal G$, the real parts of $\omega_\pm$ separate as $\mathrm{Re}\,\omega_\pm\simeq\pm\sqrt2\,\mathcal G$, while the dark pole remains fixed at $\omega_D=-i\kappa/2$. This produces the three-peak structure, with increasingly pronounced spectral minima separating the central dark contribution from the bright polariton peaks.

\section*{S5. Spectrum characterization}

The bright--dark analysis of Sec.~S4 shows that the conversion spectrum evolves from a single broad peak into a three-peak structure as the effective Raman coupling strength increases. To quantify this evolution, we characterize the spectrum using two complementary quantities: the outer full width at half maximum (FWHM) bandwidth, which measures the overall spectral width, and the internal dip depth between neighboring peaks, which quantifies the emergence of spectral splitting.

In the symmetric case (\(\kappa_\mu=\kappa_o\equiv\kappa\)) under impedance matching (\(\mathcal G_\mu=\mathcal G_o\equiv\mathcal G\)), corresponding to Fig.~2(b) of the main text, we define the dimensionless Raman coupling and FWHM bandwidth
\begin{equation}
x\equiv \frac{\mathcal G}{\kappa},
\qquad
w\equiv \frac{\Delta\omega}{\kappa}.
\tag{S72}
\end{equation}
In the lossless case, $\Gamma_r=0$, off-center local minima appear only for
$x>1/(2\sqrt{2})$.
By symmetry, the two minima occur at $\omega=\pm\omega_{\min}(x)$. Defining $\omega_{\min}(x)>0$ as the positive-frequency minimum, the corresponding minimum conversion efficiency is
\begin{equation}
\eta_{\min}(x)
\equiv
\eta\!\left[\omega_{\min}(x);x\right]
=
\frac{432x^4}
{(4x^2+1)^2(32x^2-1)},
\qquad
x>\frac{1}{2\sqrt{2}}.
\tag{S73}
\end{equation}
The dip depth relative to the unit peak efficiency is then
\begin{equation}
d(x)
\equiv
\frac{\eta_{\mathrm{peak}}-\eta_{\min}(x)}
{\eta_{\mathrm{peak}}}
=
1-\eta_{\min}(x),
\qquad
x>\frac{1}{2\sqrt{2}},
\tag{S74}
\end{equation}
where $\eta_{\mathrm{peak}}=1$ under lossless impedance-matched conditions.

The FWHM bandwidth is determined by
\begin{equation}
x^2=\frac{w(w^2+1)}{8(w+1)}.
\tag{S75}
\end{equation}
At the dip-onset point $x=1/(2\sqrt2)$, this gives
\begin{equation}
w=1,
\qquad
\Delta\omega=\kappa.
\tag{S76}
\end{equation}
For larger Raman coupling strength, the dip depth increases. Once the dip depth reaches $d=1/2$, the spectral minima fall below half of the peak value and the above-half-maximum region becomes disconnected. Beyond this point, a single FWHM bandwidth no longer provides a meaningful characterization of the spectrum.

For the finite-loss response, $\Gamma_r\neq0$, shown in Fig.~2(c) of the main text we evaluate the impedance-matched spectrum at the coupling $x=1/(2\sqrt2)$, where the lossless spectrum is maximally flat. Defining
\begin{equation}
u=\frac{\omega}{\kappa},
\qquad
\gamma=\frac{\Gamma_r}{\kappa},
\tag{S77}
\end{equation}
the conversion efficiency is
\begin{equation}
\eta(u,\gamma)
=
\frac{1/64}{
\left(u^2+\frac{1}{4}\right)
\left[
\left(u^2-\frac{1+\gamma}{4}\right)^2
+
\frac{(1+\gamma)^2}{4}u^2
\right]
}.
\tag{S78}
\end{equation}
At resonance this gives
\begin{equation}
\eta(0,\gamma)=(1+\gamma)^{-2}.
\tag{S79}
\end{equation}
The FWHM bandwidth is obtained numerically from the smallest positive solution of
\begin{equation}
\eta(u_{1/2},\gamma)=\frac{1}{2}\eta(0,\gamma),
\qquad
\Delta\omega=2\kappa u_{1/2}.
\tag{S80}
\end{equation}
The results are presented in Fig.~2(c) of the main text.

\section*{S6. Cavity-linewidth asymmetry}

We next consider the lossless case $\Gamma_r=0$ under impedance matching, while allowing the microwave and optical cavity linewidths to differ. We define the microwave-to-optical cavity-linewidth ratio
\begin{equation}
r\equiv \frac{\kappa_\mu}{\kappa_o}.
\tag{S81}
\end{equation}
Imposing the impedance-matching condition of Eq.~(S52),
$\mathcal G_\mu^2/\kappa_\mu=\mathcal G_o^2/\kappa_o$, gives
\begin{equation}
\mathcal G_\mu=\sqrt r\,\mathcal G_o .
\tag{S82}
\end{equation}

The lossless, impedance-matched conversion efficiency for arbitrary cavity-linewidth ratio $r$ is
\begin{equation}
\eta(\tilde\omega,r,\tilde {\mathcal G}_o)
=
\frac{r^2\tilde{\mathcal G}_o^4}
{
\left[
\tilde\omega^3-\tilde\omega\left(\tilde{\mathcal G}_o^2 r+\tilde{\mathcal G}_o^2+\frac{r}{4}\right)
\right]^2
+
\left[
\frac{r+1}{2}\tilde\omega^2-\tilde{\mathcal G}_o^2 r
\right]^2
},
\tag{S83}
\end{equation}
where $\tilde{\mathcal G}_o\equiv \mathcal G_o/\kappa_o$ and $\tilde\omega\equiv \omega/\kappa_o$ are dimensionless quantities defined using $\kappa_o$ as the frequency unit. 

In the $r\ll1$ limit, the microwave cavity is the narrower channel and sets
the conversion bandwidth. Although for any fixed nonzero $\tilde\omega$ the
efficiency vanishes as $r\to0$, the resonant response is resolved by introducing
the scaled detuning $\tilde\omega=r\bar\omega$. Writing $\tilde\omega=r\bar\omega$
and keeping the leading powers in $r$, Eq.~(S83) gives

\begin{equation}
\eta(\bar\omega,r,\tilde {\mathcal G}_o)
\approx
\frac{r^2 \tilde{\mathcal G}_o^4}
{
\left[-r\tilde{\mathcal G}_o^2\bar\omega+O(r^2)\right]^2
+
\left[-r\tilde{\mathcal G}_o^2+O(r^2)\right]^2
}
\approx
\frac{1}{1+\bar\omega^2}.\tag{S84}
\end{equation}

Thus, in the narrow-band limit the response approaches a Lorentzian in the
scaled detuning $\bar\omega$, with unit peak efficiency and a dimensionless
bandwidth of order $r$ in $\tilde\omega$, corresponding to an unscaled bandwidth of order $\kappa_\mu$.

Conversely, in the $r\gg1$ limit, the microwave cavity linewidth is much
broader than the optical cavity linewidth, so the conversion bandwidth is
limited primarily by $\kappa_o$. Neglecting terms of order $r^{-1}$ and
smaller in Eq.~(S83) gives

\begin{equation}
\eta(\tilde\omega,\tilde {\mathcal G}_o)
\approx
\frac{\tilde{\mathcal G}_o^4}
{
\tilde\omega^2\left(\tilde{\mathcal G}_o^2+\frac14\right)^2
+
\left(\frac{\tilde\omega^2}{2}-\tilde{\mathcal G}_o^2\right)^2
}.
\tag{S85}
\end{equation}

We evaluate the spectra at the optical Raman coupling corresponding to the dip-onset value of the symmetric lossless case,
\begin{equation}
\tilde{\mathcal G}_o=\frac{1}{2\sqrt2}.
\tag{S86}
\end{equation}
For $r=1$, this corresponds to the maximally flat spectrum of the symmetric cavity case. For $r\neq1$, impedance matching still fixes the resonant conversion efficiency at $\eta(0,r)=1$, but the spectral flatness is no longer protected by cavity symmetry [Fig.~3(a) in the main text]. At the chosen coupling $\tilde{\mathcal G}_o^2=1/8$, the conversion
efficiency in the $r\gg1$ limit, Eq.~(S85), reduces to
\begin{equation}
\eta(\tilde\omega)
\approx
\frac{1}{1+\tilde\omega^2+16\tilde\omega^4}.
\tag{S87}
\end{equation}

While the FWHM bandwidth is useful in the symmetric case, it does not always provide the most direct characterization of asymmetric spectra. We therefore define a high-efficiency bandwidth $\Delta\omega_{0.9}$ as the frequency interval for which $\eta(\omega,r)>0.9$.  Since the spectrum is even in $\omega$, we have
\begin{equation}
\Delta\omega_{0.9}=2\omega_{0.9},
\qquad
\eta(\omega_{0.9},r)=0.9,
\tag{S88}
\end{equation}
where the positive root $\omega_{0.9}>0$ is obtained numerically.

In the $r\ll1$ limit, Eq.~(S84) gives
$\Delta\omega_{0.9}=(2r/3)\kappa_o$.
Thus, the high-efficiency bandwidth, expressed in units of $\kappa_o$,
vanishes linearly with $r$. In the $r\gg1$ limit, Eq.~(S87) gives
$\Delta\omega_{0.9}/\kappa_o
=
2\sqrt{(\sqrt{73}-3)/96}
\simeq 0.48$.

The apparent asymmetry between the $r\ll1$ and $r\gg1$ limits shown in
Fig.~3(b) of the main text arises because the bandwidth is normalized to
$\kappa_o$. For $r\ll1$, the microwave cavity
is the narrower channel, so photons outside a bandwidth of order
$\kappa_\mu=r\kappa_o$ are predominantly reflected rather than converted,
giving $\Delta\omega_{0.9}/\kappa_o\propto r$. For $r\gg1$, the optical cavity
becomes the limiting channel; increasing $\kappa_\mu$ further does not broaden
the optical extraction window, so $\Delta\omega_{0.9}/\kappa_o$ approaches a
constant.

Thus, impedance matching protects the resonant conversion efficiency,
whereas cavity symmetry protects the maximally flat spectral response. As
$r$ moves away from unity, the bright--dark interference responsible for
the flat response becomes imperfect: the peak remains pinned at
$\eta(0)=1$, but the frequency window over which $\eta(\omega)$ remains
close to unity is reduced.

\section*{S7. Spatially nonuniform atom--cavity coupling}

Finally, we consider spatially nonuniform atom--cavity couplings. In Sec.~S1, the single-atom cavity couplings $g_\mu$ and $g_o$ were taken to be independent of the atomic position. Here we keep $g_\mu$ and $g_o$ as the reference coupling strengths and parametrize the position dependence by dimensionless mode functions $f_\mu(\mathbf r)$ and $f_o(\mathbf r)$:
\begin{equation}
g_{\mu,l}=g_\mu f_\mu(\mathbf r_l),
\qquad
g_{o,l}=g_o f_o(\mathbf r_l).
\tag{S89}
\end{equation}
We normalize $f_\mu$ and $f_o$ such that $f_{\mu,o}=1$ at the field maximum, so that $g_\mu$ and $g_o$ coincide with the maximum single-atom couplings. The functions $f_\mu$ and $f_o$ describe the spatial modulation of the microwave and optical couplings sampled by the atoms; in the trapped-atom geometry considered here, they correspond to the cavity-mode profiles evaluated at the atomic positions.

We assume that the classical driving fields have approximately uniform amplitudes over the atomic ensemble. In the far-detuned regime, the corresponding position-dependent effective Raman couplings, which reduce to Eq.~(S31) in the uniform-coupling limit, are given by \begin{equation} G_{\mu,l}^{(1)} \simeq -\frac{g_{\mu,l}\Omega_2}{\delta_s}, \qquad G_{o,l}^{(1)} \simeq -\frac{g_{o,l}\Omega_1}{\Delta}. \tag{S90} \end{equation} Consequently, the spatial dependence of the effective Raman couplings arises solely from the corresponding cavity--atom coupling profiles.

Under near-phase-matching conditions $\mathbf K\approx0$, the coherent atom--cavity effective interaction Hamiltonian takes the form
\begin{equation}
\frac{H_{\rm int}}{\hbar}
=
\sum_l
\left[
G_{\mu,l}^{(1)} a^\dagger \sigma_{gr}^{(l)}
+
G_{o,l}^{(1)} b^\dagger \sigma_{gr}^{(l)}
+
\mathrm{H.c.}
\right].
\tag{S91}
\end{equation}
In the weak-excitation regime, this Hamiltonian can be expressed as
\begin{equation}
\frac{H_{\rm int}}{\hbar}
=
\overline{\mathcal G}_\mu a^\dagger R_\mu
+
\overline{\mathcal G}_o b^\dagger R_o
+
\mathrm{H.c.},
\tag{S92}
\end{equation}
where, in analogy with the weighted collective mode of an inhomogeneously coupled spin ensemble in a cavity~\cite{Kurucz2011}, the microwave and optical fields address the collective spin-wave modes
\begin{equation}
R_\mu
=
\frac{1}{\overline{\mathcal G}_\mu}
\sum_l
G_{\mu,l}^{(1)}\sigma_{gr}^{(l)},
\qquad
R_o
=
\frac{1}{\overline{\mathcal G}_o}
\sum_l
G_{o,l}^{(1)}\sigma_{gr}^{(l)},
\tag{S93}
\end{equation}
with the corresponding collectively enhanced Raman couplings
\begin{equation}
\overline{\mathcal G}_\mu
=
\left(
\sum_l |G_{\mu,l}^{(1)}|^2
\right)^{1/2},
\qquad
\overline{\mathcal G}_o
=
\left(
\sum_l |G_{o,l}^{(1)}|^2
\right)^{1/2}.
\tag{S94}
\end{equation}

Spatially nonuniform couplings produce two main effects. First, they renormalize the collective Raman couplings through the root-sum-square enhancement. Compared with the uniform-coupling collective Raman couplings $\mathcal G_{\mu,o}$ defined in Eq.~(S36), the reduction factors are
\begin{equation}
\frac{\overline{\mathcal G}_{\mu,o}}{\mathcal G_{\mu,o}}
=
\left(
\frac{1}{N}
\sum_l |f_{\mu,o}(\mathbf r_l)|^2
\right)^{1/2}.
\tag{S95}
\end{equation}
For uniform couplings, $f_{\mu,o}(\mathbf r_l)=1$ for all atoms, so that $\overline{\mathcal G}_{\mu,o}=\mathcal G_{\mu,o}$ and the usual $\sqrt N$ collective enhancement is recovered.

Second, because the microwave and optical fields generally weight the atoms differently, they can address different collective spin-wave modes. The normalized spin-wave overlap is
\begin{equation}
\Lambda
=
\frac{
\sum_l
\left( G_{\mu,l}^{(1)}\right)^*
 G_{o,l}^{(1)}
}{
\left(\sum_l | G_{\mu,l}^{(1)}|^2\right)^{1/2}
\left(\sum_l | G_{o,l}^{(1)}|^2\right)^{1/2}
}.
\tag{S96}
\end{equation}
When $|\Lambda|\simeq1$, the microwave- and optical-addressed spin waves are nearly identical, and the dynamics reduce to the single-mode transducer introduced in Sec.~S2 with the replacements
$\mathcal G_\mu
\rightarrow
\overline{\mathcal G}_\mu$ and
$\mathcal G_o
\rightarrow
\overline{\mathcal G}_o$
The impedance-matching condition then becomes
\begin{equation}
\frac{\overline{\mathcal G}_\mu^2}{\kappa_\mu}
=
\frac{\overline{\mathcal G}_o^2}{\kappa_o}.
\tag{S97}
\end{equation}
For imperfect overlap, $|\Lambda|<1$, we decompose the
optical-addressed spin wave as
\begin{equation}
R_o
=
\Lambda R_\mu
+
\sqrt{1-|\Lambda|^2}\,R_\perp ,
\tag{S98}
\end{equation}
where $R_\perp$ is orthogonal to $R_\mu$. The interaction then involves four modes, $a$, $b$, $R_\mu$, and $R_\perp$. Within the independent-decay model assumed above, \(R_\mu\) and \(R_\perp\) are orthogonal collective superpositions of the same set of single-atom coherences \(\sigma_{gr}^{(l)}\), so uniform detuning and single-atom decay across the ensemble give both modes the same complex frequency \(z_r\). Solving the same
linear input-output problem as in Sec.~S3 for this four-mode system gives
\begin{equation}
\eta_{\rm swm}(\omega)
=
\frac{
\kappa_\mu\kappa_o\,
\overline{\mathcal G}_\mu^2
\overline{\mathcal G}_o^2
|\Lambda|^2
|\omega-z_r|^2
}{
|\mathcal D_\Lambda(\omega)|^2
},
\tag{S99}
\end{equation}
where
\begin{equation}
\mathcal D_\Lambda(\omega)
=
(\omega-z_r)
\left[
(\omega-z_\mu)(\omega-z_o)(\omega-z_r)
-(\omega-z_\mu)\overline{\mathcal G}_o^2
-(\omega-z_o)\overline{\mathcal G}_\mu^2
\right]
+
(1-|\Lambda|^2)
\overline{\mathcal G}_\mu^2
\overline{\mathcal G}_o^2,
\tag{S100}
\end{equation}
with $z_\mu$, $z_o$ the complex frequencies of the microwave and optical modes. This expression reduces to the single-spin-wave result Eq.~(S48) when $|\Lambda|=1$. For $|\Lambda|<1$, the overlap factor $|\Lambda|^2$
directly suppresses the conversion numerator, while the orthogonal spin-wave component, whose squared amplitude is $1-|\Lambda|^2$, introduces an additional term in $\mathcal D_\Lambda(\omega)$.

A subtle consequence of the $1-|\Lambda|^2$ term in Eq.~(S100) is that its effect is not uniformly perturbative near the spin-wave resonance. Within the homogeneous independent-decay model,
$R_\mu$ and $R_\perp$ are exactly degenerate and have the same complex
frequency $z_r$. The orthogonal component can therefore produce a
narrow coherent interference feature near the common spin-wave
resonance. In the formal lossless limit, $\Gamma_r=0$, Eq.~(S99) gives
zero conversion at $\omega=\delta_{\rm eff}$ for any $|\Lambda|<1$,
corresponding to a coherent antiresonance. A finite Rydberg-state linewidth fills in this exact zero, while away from the narrow resonant region the effect of the orthogonal spin-wave component is perturbative and scales with $1-|\Lambda|^2$. The exact degeneracy of the two spin-wave modes is specific to the homogeneous independent-decay approximation and can be lifted by mode-dependent shifts or collective decay.

For a representative cavity geometry, we next estimate the magnitude of
the spin-wave mismatch. We take the microwave coupling to be approximately uniform over the ensemble, $f_\mu(\mathbf r_\perp)\simeq1$, assuming that the atoms are trapped near a microwave-field maximum at approximately fixed distance from the chip surface~\cite{Tsiamis2026}. By contrast, for the optical cavity we consider a running-wave Gaussian mode. We assume that the atomic ensemble is centered near the optical cavity-mode waist and that its longitudinal extent is much smaller than the Rayleigh range of the optical cavity mode. Diffraction-induced variations of the transverse mode size and Gouy phase across the ensemble can therefore be neglected. The remaining longitudinal propagation phase is absorbed into the spin-wave phase-matching factor, while the spatial variation of the optical coupling magnitude is determined by the transverse Gaussian profile
\begin{equation}
f_o(\mathbf r_\perp)
=
e^{-r_\perp^2/w_o^2},
\tag{S101}
\end{equation}
where $\mathbf r_\perp$ is measured relative to the optical cavity axis, $r_\perp=|\mathbf r_\perp|$, and $w_o$ is the optical cavity waist. For a Gaussian transverse atomic density distribution with equal rms widths in the two transverse directions, $\langle x^2\rangle=\langle y^2\rangle=\sigma_\perp^2$, Eq.~(S95) gives
\begin{equation}
\frac{\overline{\mathcal G}_\mu}
{\mathcal G_\mu}
\simeq
1,
\qquad
\frac{\overline{\mathcal G}_o}
{\mathcal G_o}
=
\frac{1}
{\sqrt{1+4\sigma_\perp^2/w_o^2}}.
\tag{S102}
\end{equation}
Thus, in this geometry the microwave collective coupling is essentially unchanged, while the optical collective coupling is reduced because atoms away from the cavity axis sample a weaker optical field. Evaluating Eq.~(S96) for the same geometry gives
\begin{equation}
|\Lambda|^2
=
\frac{
1+4\sigma_\perp^2/w_o^2
}{
\left(1+2\sigma_\perp^2/w_o^2\right)^2
}.
\tag{S103}
\end{equation}
Figure~3(c) of the main text shows both $\overline{\mathcal G}_o/\mathcal G_o$ and $|\Lambda|^2$ as functions of $\sigma_\perp/w_o$. In the limit $\sigma_\perp\ll w_o$, the optical collective coupling approaches its uniform-coupling value and the spin-wave overlap approaches unity. Conversely, for clouds comparable to or larger than the optical waist, spatial nonuniformity both reduces the optical collective enhancement and introduces a mismatch between the microwave- and optical-addressed spin waves. 

For comparison with an experimentally demonstrated confinement,
Ref.~\cite{Greve2022} reports a transverse rms cloud radius
$r_{\rm rms}=4.7(8)\,\mu{\rm m}$ in an optical cavity with waist
$w_o=72\,\mu{\rm m}$. For the isotropic Gaussian distribution used
here, $r_{\rm rms}=\sqrt{\langle r_\perp^2\rangle}
=\sqrt{2}\sigma_\perp$, so that
\begin{equation}
\frac{\sigma_\perp}{w_o}
=
\frac{r_{\rm rms}}{\sqrt{2}\,w_o}
\simeq 0.046.
\tag{S104}
\end{equation}
Substituting this value into Eq.~(S103) gives
$|\Lambda|^2\simeq0.99998$, corresponding to a spin-wave mismatch
$1-|\Lambda|^2\simeq1.8\times10^{-5}$.

\begin{figure}[t]
\centering
\includegraphics[width=0.82\linewidth]
{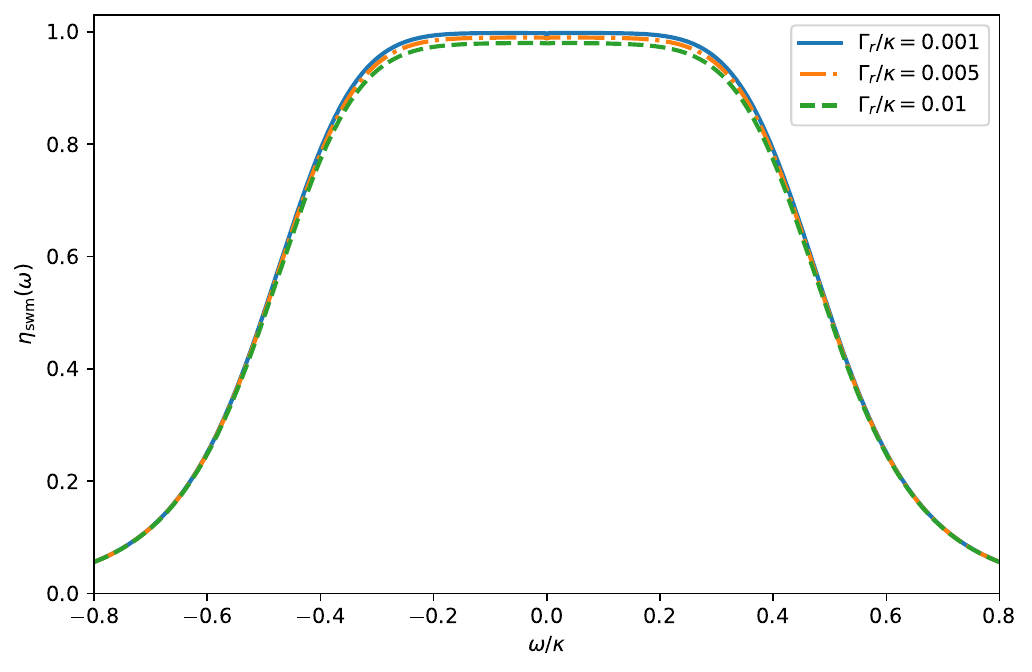}
\caption{
Full two-spin-wave conversion spectrum from Eq.~(S99) for the spin-wave
overlap $|\Lambda|^2=0.99998$ corresponding to the confinement reported
in Ref.~\cite{Greve2022}.
The curves correspond to different Rydberg-state linewidths,
$\Gamma_r/\kappa=0.001$ (solid), $0.005$ (dash-dotted), and $0.01$
(dashed). The remaining parameters are
$\kappa_\mu=\kappa_o\equiv\kappa$,
$\overline{\mathcal G}_\mu=\overline{\mathcal G}_o
=\kappa/(2\sqrt{2})$, and $\delta_{\rm eff}=0$.
}
\label{fig:swm_spectrum}
\end{figure}

To illustrate the effect of this small mismatch on the full two-spin-wave conversion spectrum, Fig.~\ref{fig:swm_spectrum} shows the conversion spectrum from Eq.~(S99) for
$|\Lambda|^2=0.99998$ and several finite Rydberg-state linewidths.
We take $\kappa_\mu=\kappa_o\equiv\kappa$,
$\overline{\mathcal G}_\mu=\overline{\mathcal G}_o=\kappa/(2\sqrt{2})$,
and $\delta_{\rm eff}=0$. For the finite linewidths shown, the small
spin-wave mismatch produces only a small modification of the conversion
spectrum.

\section*{S8. Compatibility of strong effective coupling with adiabatic elimination}

The broadband conversion regime analyzed in Secs.~S3 and S4 is reached when the
effective Raman couplings become comparable to the cavity linewidths,
$\lvert\mathcal{G}_{o,\mu}\rvert\sim\kappa_{o,\mu}$. This does not conflict
with adiabatic elimination of the intermediate atomic states. The reason is
that the adiabatic-elimination conditions constrain the admixture of the
eliminated states, whereas $\mathcal{G}_o$ and $\mathcal{G}_\mu$ describe
induced couplings within the reduced slow manifold.

To make this distinction explicit, we introduce the small dispersive parameters
\begin{equation}
\epsilon_o
=
\frac{\sqrt{N}\,\lvert g_o\rvert}{\lvert\Delta\rvert},
\qquad
\xi_o
=
\frac{\lvert\Omega_1\rvert}{\lvert\Delta\rvert},
\qquad
\epsilon_\mu
=
\frac{\sqrt{N}\,\lvert\Omega_2\rvert}{\lvert\delta_s\rvert},
\qquad
\xi_\mu
=
\frac{\lvert g_\mu\rvert}{\lvert\delta_s\rvert}.\tag{S105}
\end{equation}
For low intracavity occupation, the dispersive regime corresponds to
$\epsilon_o,\xi_o,\epsilon_\mu,\xi_\mu\ll1$. In this limit, the eliminated
states remain only weakly populated, with corrections scaling as
$\mathcal{O}(\epsilon_o^2,\xi_o^2,\epsilon_\mu^2,\xi_\mu^2)$. At the same time,
the effective Raman couplings in the slow manifold scale as
\begin{equation}
\lvert\mathcal{G}_o\rvert
\simeq
\frac{
\sqrt{N}\,\lvert g_o\rvert\,\lvert\Omega_1\rvert
}{
\lvert\Delta\rvert
}
=
\epsilon_o\xi_o\lvert\Delta\rvert,
\qquad
\lvert\mathcal{G}_\mu\rvert
\simeq
\frac{
\lvert g_\mu\rvert\sqrt{N}\,\lvert\Omega_2\rvert
}{
\lvert\delta_s\rvert
}
=
\epsilon_\mu\xi_\mu\lvert\delta_s\rvert.
\tag{S106}
\end{equation}
Thus, even when each admixture parameter is small, the products
$\epsilon_o\xi_o\lvert\Delta\rvert$ and
$\epsilon_\mu\xi_\mu\lvert\delta_s\rvert$ can be comparable to the cavity
linewidths for sufficiently large detunings and correspondingly large bare
couplings and Rabi frequencies.

For the optical Raman coupling, the same scaling can be written as
\begin{equation}
\lvert\mathcal{G}_o\rvert
\simeq
\frac{
\sqrt{N}\,\lvert g_o\rvert\,\lvert\Omega_1\rvert
}{
\lvert\Delta\rvert
}
=
\xi_o\sqrt{N}\,\lvert g_o\rvert
=
\epsilon_o\lvert\Omega_1\rvert.
\tag{S107}
\end{equation}
Combining this identity with
$\lvert\mathcal{G}_o\rvert\sim\kappa_o$ and
$\epsilon_o,\xi_o\ll1$ gives
\begin{equation}
\sqrt{N}\,\lvert g_o\rvert
\sim
\frac{\kappa_o}{\xi_o}
\gg
\kappa_o,
\qquad
\lvert\Omega_1\rvert
\sim
\frac{\kappa_o}{\epsilon_o}
\gg
\kappa_o.
\tag{S108}
\end{equation}
Together with the dispersive requirements
$\sqrt{N}\,\lvert g_o\rvert,\lvert\Omega_1\rvert
\ll\lvert\Delta\rvert$, this gives the optical hierarchy
\begin{equation}
\kappa_o
\ll
\sqrt{N}\,\lvert g_o\rvert,\ \lvert\Omega_1\rvert
\ll
\lvert\Delta\rvert.
\label{eq:optical-hierarchy}
\tag{S109}
\end{equation}

Similarly, for the microwave Raman coupling one may write
\begin{equation}
\lvert\mathcal{G}_\mu\rvert
\simeq
\frac{
\lvert g_\mu\rvert\sqrt{N}\,\lvert\Omega_2\rvert
}{
\lvert\delta_s\rvert
}
=
\xi_\mu\sqrt{N}\,\lvert\Omega_2\rvert
=
\epsilon_\mu\lvert g_\mu\rvert.
\tag{S110}
\end{equation}
Combining this identity with
$\lvert\mathcal{G}_\mu\rvert\sim\kappa_\mu$ and
$\epsilon_\mu,\xi_\mu\ll1$ gives
\begin{equation}
\sqrt{N}\,\lvert\Omega_2\rvert
\sim
\frac{\kappa_\mu}{\xi_\mu}
\gg
\kappa_\mu,
\qquad
\lvert g_\mu\rvert
\sim
\frac{\kappa_\mu}{\epsilon_\mu}
\gg
\kappa_\mu.
\tag{S111}
\end{equation}
Together with the dispersive requirements
$\lvert g_\mu\rvert,\sqrt{N}\,\lvert\Omega_2\rvert
\ll\lvert\delta_s\rvert$, this gives the microwave hierarchy
\begin{equation}
\kappa_\mu
\ll
\lvert g_\mu\rvert,\ \sqrt{N}\,\lvert\Omega_2\rvert
\ll
\lvert\delta_s\rvert.
\label{eq:microwave-hierarchy}
\tag{S112}
\end{equation}

The inequalities in Eqs.~(S109)
and~(S112) are not additional assumptions required by
adiabatic elimination itself. Rather, their upper bounds are set by the
adiabatic-elimination conditions, while the lower bounds relative to
$\kappa_{o,\mu}$ arise from additionally imposing the strong-hybridization
condition
$\lvert\mathcal{G}_{o,\mu}\rvert\sim\kappa_{o,\mu}$. Away from this regime,
when
$\lvert\mathcal{G}_{o,\mu}\rvert/\kappa_{o,\mu}<1$, the lower scale entering
the corresponding hierarchy is
$\lvert\mathcal{G}_{o,\mu}\rvert$ rather than $\kappa_{o,\mu}$. The
lower-bound requirements on the bare couplings and Rabi
frequencies are therefore correspondingly relaxed, thereby widening the
parameter window compatible with both adiabatic elimination and the desired
operating point. Thus, the intermediate states can remain perturbatively
eliminated while the reduced modes $(a,b,R)$ enter the strong-hybridization
regime relevant for broadband impedance-matched conversion.

For a given experimental setup, however, the bare atom--cavity couplings $g_o$ and $g_\mu$, the cavity linewidths
$\kappa_o$ and $\kappa_\mu$, and the atom number $N$ are fixed device
parameters. Nevertheless, the effective Raman couplings can be tuned
\textit{in situ} by changing the ratios $\Omega_1/\Delta$ and
$\Omega_2/\delta_s$. This tuning may be implemented by varying the driving-field Rabi frequencies at fixed detunings, as considered below, by varying the detunings at fixed Rabi frequencies, or by jointly adjusting both. In all cases, the parameters must continue to satisfy
\begin{equation}
\lvert\Delta\rvert
\gg
\sqrt{N}\,\lvert g_o\rvert,\ \lvert\Omega_1\rvert,
\qquad
\lvert\delta_s\rvert
\gg
\lvert g_\mu\rvert,\ \sqrt{N}\,\lvert\Omega_2\rvert.
\tag{S113}
\end{equation}

\paragraph{Representative tunable parameter range.---}
To illustrate a representative experimentally accessible regime, we consider
parameters motivated by the platform of trapped Rydberg atoms coupled to a
superconducting coplanar-waveguide (CPW) resonator discussed in the main text.
In particular, the single-atom microwave coupling
$\lvert g_\mu\rvert/2\pi=10~\mathrm{MHz}$ is estimated in
Ref.~\cite{Tsiamis2026} for Rydberg atoms trapped about
$10~\mu\mathrm{m}$ from the superconducting chip surface, based on the
experimentally characterized Rydberg--CPW platform and cavity-field
modeling of Ref.~\cite{Kaiser2022}. For the optical coupling, we take
$\lvert g_o\rvert/2\pi=0.5~\mathrm{MHz}$, consistent with
Ref.~\cite{Greve2022}. Taking $N=400$ then gives
$\sqrt{N}\lvert g_o\rvert/2\pi\simeq10~\mathrm{MHz}$. The remaining parameters are chosen such that the effective Raman couplings
can approach the cavity linewidths while the intermediate states remain
perturbatively eliminated. Specifically, we take equal optical and microwave
cavity linewidths, $\kappa_o=\kappa_\mu\equiv\kappa$, and choose the fixed
device parameters
\begin{equation}
\frac{\kappa_o}{2\pi}
=
\frac{\kappa_\mu}{2\pi}
=
1~\mathrm{MHz},
\qquad
\frac{\sqrt{N}\,\lvert g_o\rvert}{2\pi}
=
\frac{\lvert g_\mu\rvert}{2\pi}
=
10~\mathrm{MHz},
\qquad
N=400.
\label{eq:representative-fixed-parameters}
\tag{S114}
\end{equation}
We further take
\begin{equation}
\frac{\lvert\Delta\rvert}{2\pi}
=
\frac{\lvert\delta_s\rvert}{2\pi}
=
100~\mathrm{MHz}.
\tag{S115}
\label{eq:representative-detunings}
\end{equation}
The corresponding fixed dispersive ratios are
\begin{equation}
\frac{\sqrt{N}\,\lvert g_o\rvert}{\lvert\Delta\rvert}
=
\frac{\lvert g_\mu\rvert}{\lvert\delta_s\rvert}
=
0.1.
\label{eq:representative-fixed-ratios}
\tag{S116}
\end{equation}

Under impedance-matching conditions, $|\mathcal G_o|=|\mathcal G_\mu|$, we parameterize the common effective coupling as
\begin{equation}
|\mathcal G_o|=|\mathcal G_\mu|=x\kappa,
\label{eq}
\tag{S117}
\end{equation}
where $x$ is the dimensionless Raman coupling introduced in Eq.~(S72).

For the fixed parameters in Eqs.~(S114) and (S115), the driving-field Rabi frequencies required to satisfy the impedance-matching condition are
\begin{equation}
\frac{\lvert\Omega_1\rvert}{2\pi}
=
10x~\mathrm{MHz},
\qquad
\frac{\sqrt{N}\,\lvert\Omega_2\rvert}{2\pi}
=
10x~\mathrm{MHz}.
\tag{S118}
\label{eq:representative-control-fields}
\end{equation}
For $N=400$, the latter relation reduces to
\begin{equation}
\frac{\lvert\Omega_2\rvert}{2\pi}
=
0.5x~\mathrm{MHz}
\tag{S119}.
\end{equation}
The corresponding drive-dependent dispersive ratios are
\begin{equation}
\frac{\lvert\Omega_1\rvert}{\lvert\Delta\rvert}
=
\frac{\sqrt{N}\,\lvert\Omega_2\rvert}
{\lvert\delta_s\rvert}
=
0.1x.
\label{eq:representative-drive-ratios}
\tag{S120}
\end{equation}
The conversion bandwidth follows from Eq.~(S75). Defining the normalized bandwidth as $w=\Delta\omega/\kappa$, the relation between $w$ and $x$ is
\begin{equation}
x^2
=
\frac{w(w^2+1)}{8(w+1)}.
\label{eq:representative-bandwidth}
\tag{S121}
\end{equation}

Thus, tuning $x$ from $0.01$ to $1$ changes the effective Raman coupling
from $0.01\kappa$ to $\kappa$ and broadens the conversion bandwidth from
approximately $0.8~\mathrm{kHz}$ to $3.1~\mathrm{MHz}$. Over the same
interval, $\lvert\Omega_1\rvert/(2\pi)$ increases from
$0.1~\mathrm{MHz}$ to $10~\mathrm{MHz}$, while
$\lvert\Omega_2\rvert/(2\pi)$ increases from $0.005~\mathrm{MHz}$ to
$0.5~\mathrm{MHz}$. The corresponding Rabi-frequency-to-detuning ratios
increase from $0.001$ to $0.1$, whereas the cavity linewidths, bare
atom--cavity couplings, atom number, and detunings remain fixed. At the
upper-coupling endpoint, $x=1$, the largest Rabi-frequency-to-detuning
ratio is $0.1$, so the system remains in the dispersive regime.

Representative operating points are summarized in the Table

\begin{equation}
\begin{array}{c|c|c|c|c|c}
x
&
\lvert\mathcal{G}_{o,\mu}\rvert/(2\pi)
&
\lvert\Omega_1\rvert/(2\pi)
&
\lvert\Omega_2\rvert/(2\pi)
&
\lvert\Omega_1\rvert/\lvert\Delta\rvert
&
\Delta\omega/(2\pi)
\\
\hline
1
&
1~\mathrm{MHz}
&
10~\mathrm{MHz}
&
0.5~\mathrm{MHz}
&
0.1
&
3.10~\mathrm{MHz}
\\
0.1
&
0.1~\mathrm{MHz}
&
1~\mathrm{MHz}
&
0.05~\mathrm{MHz}
&
0.01
&
86.3~\mathrm{kHz}
\\
0.05
&
0.05~\mathrm{MHz}
&
0.5~\mathrm{MHz}
&
0.025~\mathrm{MHz}
&
0.005
&
20.4~\mathrm{kHz}
\\
0.01
&
0.01~\mathrm{MHz}
&
0.1~\mathrm{MHz}
&
0.005~\mathrm{MHz}
&
0.001
&
0.801~\mathrm{kHz}
\end{array}
\tag{S122}
\label{eq:tuning-table}
\end{equation}
Here the Rabi-frequency-to-detuning column also equals
$\sqrt{N}\lvert\Omega_2\rvert/\lvert\delta_s\rvert$.

As shown in Table~(S122), the same fixed device spans more than three orders of magnitude in conversion bandwidth through \textit{in situ} control of the driving fields. At the lower-coupling endpoint, the narrow bandwidth enables spectrally selective conversion, whereas the upper-coupling endpoint produces a megahertz-scale broadband response. The above-half-maximum region remains connected across the entire tuning range. Moreover, the drive-to-detuning ratios remain at or below $0.1$ and decrease with decreasing $x$. Consequently, the driving-field contributions to the adiabatic-elimination conditions become progressively better satisfied as $x$ is reduced.

The representative values above apply in the low-excitation regime.
Appreciable intracavity occupations tighten the adiabatic-elimination
conditions through the factors
$\sqrt{\langle a^\dagger a\rangle+1}$ and
$\sqrt{\langle b^\dagger b\rangle+1}$ and may therefore require larger
detunings.

\end{document}